\documentclass[showpacs,prl,onecolumn,aps,superscriptaddress,preprintnumbers,nofootinbib]{revtex4}

\usepackage[T1]{fontenc}
\usepackage[latin9]{inputenc}
\usepackage{amsmath,amssymb}
\usepackage{epsfig}
\usepackage{graphicx}
\usepackage{amsmath}
\usepackage{amsfonts}
\usepackage{epstopdf}
\def\slashchar#1{\setbox0=\hbox{$#1$}     		
   \dimen0=\wd0                                 	
   \setbox1=\hbox{/} \dimen1=\wd1               	
   \ifdim\dimen0>\dimen1                        	
      \rlap{\hbox to \dimen0{\hfil/\hfil}}      	
      #1                                        	
   \else                                        	
      \rlap{\hbox to \dimen1{\hfil$#1$\hfil}}   	
      /                                         	
   \fi}

\renewcommand{\vec}{\boldsymbol}
\newcommand{\beq}{\begin{equation}}
\newcommand{\eeq}{\end{equation}}
\newcommand{\bea}{\begin{eqnarray}}
\newcommand{\eea}{\end{eqnarray}}
\newcommand{\ba}{\begin{array}}
\newcommand{\ea}{\end{array}}

\def\eq#1{{Eq.~(\ref{#1})}}
\def\fig#1{{Fig.~\ref{#1}}}
\newcommand{\bas}{\bar{\alpha}_S}

\newcommand{\nn}{\nonumber}

\newcommand{\h}{\frac{1}{2}}

\newcommand{\ha}{{\cal H}}

\newcommand{\Lb}{\left(}
\newcommand{\Rb}{\right)}
\def\sp{\langle S^2\rangle}
\def\pom{{I\!\!P}}

\begin{document}

\title{ A new parton model for the soft interactions at high energies: two channel approximation.	}
\author{E. ~Gotsman}
\email{gotsman@post.tau.ac.il}
\affiliation{Department of Particle Physics, School of Physics and Astronomy,
Raymond and Beverly Sackler
 Faculty of Exact Science, Tel Aviv University, Tel Aviv, 69978, Israel}
 \author{ E.~ Levin}
\email{leving@tauex.tau.ac.il, eugeny.levin@usm.cl}
\affiliation{Department of Particle Physics, School of Physics and Astronomy,
Raymond and Beverly Sackler
 Faculty of Exact Science, Tel Aviv University, Tel Aviv, 69978, Israel}
 \affiliation{Departemento de F\'isica, Universidad T\'ecnica Federico
 Santa Mar\'ia, and Centro Cient\'ifico-\\
Tecnol\'ogico de Valpara\'iso, Avda. Espana 1680, Casilla 110-V,
 Valpara\'iso, Chile} 
 \author{  I.~ Potashnikova}
\email{irina.potashnikova@usm.cl}
\affiliation{Departemento de F\'isica, Universidad T\'ecnica Federico
 Santa Mar\'ia, and Centro Cient\'ifico-\\
Tecnol\'ogico de Valpara\'iso, Avda. Espana 1680, Casilla 110-V,
 Valpara\'iso, Chile}
\date{\today}

\keywords{BFKL Pomeron, soft interaction, CGC/saturation approach, correlations}
\pacs{ 12.38.-t,24.85.+p,25.75.-q}

\begin{abstract}

The primary goal of this paper is  to describe the diffraction production
 using the  model that
takes into account the Pomeron interaction, and satisfies both $t$ and $s$
 channel unitarity. We hope that 
these features will allow us to describe the diffraction production in
 a more convenient way than in CGC motivated models, that do not 
satisfy these
 unitarity constraints. Unfortunately,  we show that both approaches are
only able to describe  half of the cross section for the single
 diffraction production, leaving the second half to be estimates
of  the large mass production in the Pomeron approach.
 The impact parameter 
dependance
 of the scattering amplitudes show that  soft interactions at high
 energies measured at the LHC,  have a much richer structure than   
  presumed.  We discuss the $t$-dependence of the elastic cross section in
 wide range  of $|t|=0 \div 1 \,GeV^2 $. We show that in the kinematic
 region of the minimum,  we cannot use  approximate formulae to 
calculate
 the real part of the amplitude. The exact calculation in our model, shows
 that the real part is rather small, and it is necessary  to  
 include  the Odderon
 contribution  in order  to describe the experimental data.

 \end{abstract}
 
 \preprint{TAUP-}

\maketitle

\tableofcontents


 \section{ Introduction}
 
  In our recent  paper \cite{GLPPM} we  proposed 
a new parton model  for high energy soft interactions, which is based on 
 Pomeron calculus in 1+1 space-time  dimensions, suggested in Ref. 
\cite{KLL}, and on
 simple assumptions of hadron structure, related to the impact 
parameter
 dependence of the scattering amplitude. This parton model stems from QCD,
 assuming that the unknown non-perturbative corrections lead to 
determining  the
 size of the interacting dipoles. The advantage of this approach is that 
it
 satisfies both the $t$-channel and  $s$-channel unitarity, and can be 
used
 for summing all diagrams of the Pomeron interaction including  Pomeron
 loops. Hence,  we can use this approach for all possible
 reactions: dilute-dilute (hadron-hadron), dilute-dense (hadron-nucleus)
 and dense-dense (nucleus-nucleus) parton system scattering.  
 
 In other words, in this model we assume that the dimensional scale,
 that determines the interaction at high energy, arises from the
 non-perturbative QCD approach,  which fixes  the  size of  dipoles.
 Such an approach is quite different from the Colour Glass Condensate
 (CGC)  one,  where  this scale  originates from the interaction of
 dipoles at short distances,  and turns out to be large and increases 
with
  energy\cite{KOLEB}. In spite of the fact, that the model, based on CGC
 approach\cite{GLP1,GLP2,GLMNI,GLM2CH,GLMINCL,GLMCOR,GLMSP,GLMACOR},
 describes all available data on soft interactions at high energy as
 well as the deep inelastic processes, it has an intrinsic problem:
 the CGC approach, in it's  present form, does not provide a scattering
 amplitude, that satisfies both the $t$ and $s$ channel 
unitarity\cite{KLL}.
 
 We have shown  that the new parton  model is able to describe high
 energy data on the
 total and  elastic cross sections for proton-proton scattering, but 
the simple version of
 Ref.\cite{GLPPM}  leads to vanishing of diffractive production. In
 this paper we propose  a two channel model which generates 
 diffraction production in the region of small masses.  As it
 is well known from the experiences (see for example
 Refs.\cite{GLM2CH,KMR} ) that  diffraction production is the
 process, which is  difficult to describe and, specifically, this
 process  provides a check of our approach to  long distances
 physics, which is  part of non-perturbative QCD.

 We  demonstrate that a two channel model is able to describe  
four 
experimental
 observables: $\sigma_{\rm tot}$,$\sigma_{\rm el}$,  $B_{\rm el}$ and
 the single diffraction cross sections. We show that this model
 leads to  a rich  structure for the impact parameter dependence of 
the
 scattering amplitude. In particular, we study the dependence of the
 elastic cross section as function of $|t| = 0 \div 1\,GeV^2$. We show
  that we are able to describe the experimental data on
 $d \sigma_{el}/d t$ in this $t$-region: the position of the
 minimum with $|t|_{min} =0.52\,GeV^2$ at $W=7 \,TeV$ and the
 value and $t$ behaviour for larger $t$.

\section{ The new parton model} 

\subsection{General approach.}
  As we have discussed
 in Ref.\cite{GLPPM,KLL} the new parton model is  based on three
 ingredients:
 
 1. The Colour Glass Condensate
(GCC) approach (see Ref.\cite{KOLEB} for a review), which can be
 re-written in the equivalent form as the interaction of  BFKL
 Pomerons\cite{AKLL} in a limited range of rapidities
 ( $Y \leq Y_{\rm max}$):

 \beq \label{RAPRA}
Y \,\leq\,\frac{2}{\Delta_{\mbox{\tiny BFKL}}}\,\ln\Lb
 \frac{1}{\Delta^2_{\mbox
{\tiny BFKL}}}\Rb
\eeq 
 $\Delta_{\mbox{\tiny BFKL}}$ denotes the intercept of the BFKL 
 Pomeron\cite{BFKL}. In our model $ \Delta_{\mbox{\tiny BFKL}}\,
\approx\,0.2 - 0.25$    leading to $Y_{max} = 20 - 30$, which covers
 all collider energies.     

2. The   following Hamiltonian:
  \begin{equation}\label{HNPM}
\ha_{\rm NPM}=-\frac{1}{\gamma}\bar PP\eeq
where NPM stands for ``new parton model''. $P$ and $\bar P$ are the BFKL
 Pomeron fields.
The fact that it is self dual is evident.  This Hamiltonian in the limit
 of small $\bar P$ reproduces the  Balitsky-Kovchegov Hamiltonian
 $\ha_{\rm BK}$
( see Ref.\cite{KLL} for details). This condition is the most
 important one for  determining the form of
$\ha_{\rm NPM}$.  $\gamma$ in \eq{HNPM} denotes the dipole-dipole
 scattering amplitude, which in QCD is proportional to $\bas^2$.
 
 3. The new commutation relations:
\beq\label{CRCOR}
\Big(1\,\,-\,\,P\Big)\Big(1\,\,-\,\,\bar P \Big)\,\,=\,\,(1-\gamma)\Big(1\,\,-\,\,\bar P\Big) \Big(1\,\,-\,\,P\Big)
\eeq
For small $\gamma$  and in the regime where  $P$ and $\bar P$ are also 
small, we obtain
\beq
[P,\bar P]=-\gamma +...
\eeq
consistent with the standard  BFKL Pomeron calculus (see Ref.\cite{KLL}
 for details) . 

 In Ref.\cite{KLL}, it was proved that the scattering matrix 
for the model 
is
  given by 
\begin{eqnarray}\label{classs}
S^{\rm NPM}_{m\bar n}(Y)&=&e^{\frac{1}{\gamma} \int_0^Yd\eta\left[
 \ln(1-p)\frac{\partial}{\partial \eta}\ln (1-\bar p) 
+\bar pp\right]}[1-p(Y)]^m[1-\bar p(0)]^{\bar n}|_{p(0)=1-e^{-\gamma
 \bar n};\  \bar p(Y)=1-e^{-\gamma m}}\nonumber\\
&=&[1-p(Y)]^m\,e^{\frac{1}{\gamma}\int_0^Yd\eta \left[\ln(1-\bar p)+\bar
 p\right]p}
\end{eqnarray}
where $p(\eta)$ and $\bar p(\eta)$ are solutions of the classical equations
 of motion and have the form:

 \beq \label{H03}
 P (\eta)\,=\,\frac{ \alpha +\beta e^{ (1 - \alpha) \eta} }{1 + \beta e^{ (1
 - \alpha)  \eta}}; \ \ \ \ \bar P(\eta)=   \frac{ \alpha (1+\beta e^{
 (1 - \alpha) \eta}) }{\alpha +  \beta e^{ (1 - \alpha)  \eta}};
 \eeq
 where the parameters $\beta$ and $\alpha$ should be  determined from 
the
 boundary conditions:
 \beq \label{H0BC}
 P (\eta= 0)\,=\,p_0;\,\,\,\,\,\,\,\, \bar P (\eta= Y)\,=\,\frac{\alpha}{P
 (\eta= Y)}\,=\,\bar p_0
 \eeq

It is interesting to compare the scattering amplitude given by this
 expression to that obtained from the BK equation, which  describes
  deep inelastic scattering  on   nuclei in QCD. For  which  
we have
\beq
S^{\rm BK}_{m\bar n}(Y)=\int dP(\eta)d\bar P(\eta)e^{\frac{1}{\gamma}
 \int_0^Yd\eta\left[ \ln(1-P)\frac{\partial}{\partial \eta}\ln (1-\bar P) -
\ln (1-\bar P)PP\right]}(1-P(Y))^m(1-\bar P(0))^{\bar n}
\eeq
In the classical approximation
\begin{eqnarray}\label{classs4}
S^{\rm BK}_{m\bar n}(Y)&=&e^{\frac{1}{\gamma} \int_0^Yd\eta\left[
 \ln(1-p)\frac{\partial}{\partial \eta}\ln (1-\bar p) 
-\ln(1-\bar p)p\right]}[1-p(Y)]^m[1-\bar p(0)]^{\bar n}|_{p(0)=1-e^{-\gamma
 \bar n};\  \bar p(Y)=1-e^{-\gamma m}}\nonumber\\
&=&[1-p(Y)]^m
\end{eqnarray}
Note, that the solution for $\bar P$, is not relevant for the BK 
amplitude,
 which is determined entirely by $P(Y)$. On the other hand, the scattering
 amplitude in the NPM  depends on $\bar P$. Nevertheless, the two 
models
 should be  similar  in the regime where the  BK evolution  is 
valid. The
 results of the estimates in Ref.\cite{KLL} shows that  in the region
  close to  saturation, the differences between BK and NPM are 
quite significant.

 \subsection{ Interrelation with QCD.} 
As  has been mentioned, in the limited range of energies, given by
 \eq {RAPRA}, both QCD and our model describe the interaction of the
 BFKL Pomerons\cite{BFKL}. For  weak fields $P$ and $\bar P$, the 
model
 reproduces the BK limit of the CGC approach,  assuming that the
 non-perturbative corrections result   in  determining the size of
 the interacting dipoles, and hence, the successful description of the 
soft
 data at high energies in CGC approach
\cite{GLP1,GLP2,GLMNI,GLM2CH,GLMINCL,GLMCOR,GLMSP,GLMACOR} supports
 the idea that this effective size is rather small. The model leads
 to the descriptions that satisfy both $t$-unitarity and
 $s$-channel unitarity, while, as it was shown in Ref.\cite{KLL},
 the BFKL Pomeron calculus in the BK limit, as well as the Braun
 Hamiltonian\cite{BRAUN}  for dense-dense system scattering
  violates $s$-channel unitarity. Unfortunately, we are still
 far from being able to solve  this problem in the effective
 QCD theory at high energy (i.e. in the CGC /saturation approach).

 \subsection{ Two channel approximation } 
 Our model includes three essential ingredients: (i) the new parton
 model for the dipole-dipole scattering amplitude that has been discussed
 above; (ii) the simplified two channel model that  enables  us to 
take
 into account  diffractive production in  the low mass region, and (iii)
 the assumptions for impact parameter dependence of the initial conditions.
 
 In the two channel approximation we replace the rich structure of the
 diffractively produced  states, by a single  state with the wave
 function $\psi_D$.
  The observed physical 
hadronic and diffractive states are written in the form 
\beq \label{MF1}
\psi_h\,=\,\alpha\,\Psi_1+\beta\,\Psi_2\,;\,\,\,\,\,\,\,\,\,\,
\psi_D\,=\,-\beta\,\Psi_1+\alpha \,\Psi_2;~~~~~~~~~
\mbox{where}~~~~~~~ \alpha^2+\beta^2\,=\,1;
\eeq 

Functions $\psi_1$ and $\psi_2$  form a  
complete set of orthogonal
functions $\{ \psi_i \}$ which diagonalize the
interaction matrix ${\bf T}$
\beq \label{GT1}
A^{i'k'}_{i,k}=<\psi_i\,\psi_k|\mathbf{T}|\psi_{i'}\,\psi_{k'}>=
A_{i,k}\,\delta_{i,i'}\,\delta_{k,k'}.
\eeq
The unitarity constraints take  the form
\beq \label{UNIT}
2\,\mbox{Im}\,A_{i,k}\left(s,b\right)=|A_{i,k}\left(s,b\right)|^2
+G^{in}_{i,k}(s,b),
\eeq
where $G^{in}_{i,k}$ denotes the contribution of all non 
diffractive inelastic processes,
i.e. it is the summed probability for these final states to be
produced in the scattering of a state $i$ off a state $k$. In \eq{UNIT} 
$\sqrt{s}  = W$ denotes the energy of the colliding hadrons and $b$ 
denotes 
the 
impact  parameter. In our approach we used the solution to  \eq{UNIT}
given by  \eq{classs} and 
\beq \label{AIK}
A_{ik} \,=\,1 - S^{\rm NPM}_{i k}(Y)
\eeq

  \subsection{ The general formulae.} 

 {\it  Initial conditions:}
 Following Ref.\cite{GLPPM} we chose the initial conditions in the form:
\begin{equation} \label{IC}
p_i(b') = p_{0 i} \,S(b',m_i)~~~\mbox{with}~~S(b,m_i)= m_i b K_1(m_i b );~~~~~
\bar{p_i}(\vec{b} - \vec{b}) = p_{0i} S( \vec{b} - \vec{b}',m_i) ~~~~~~~z_m = e^{\Delta(1 - p_{01})Y}
\end{equation}

Both $p_{0i} $ and masses $m_i$, as well as the Pomeron intercept
 $\Delta$, are  parameters of the model,  which are determined by 
fitting to the relevant data. Note, that
 $S\Lb b, m_i\Rb \xrightarrow{m_i\,b \gg 1}\,\exp\Lb - m_i\,b\Rb$
 in accord with the Froissart theorem\cite{FROI},

From \eq{IC}  we find that

\bea 
a_{ik}(b,b')\,\equiv\, a_{i,k}\Lb p_i, \bar{p}_k, z_m\Rb &=&\,\h\Lb p_{i} + \bar{p}_{k}\Rb  \,+\,\frac{1}{2\,z_m}\Lb (1-p_i)(1-\bar{p}_k ) \,-\, D_{i,k}\Rb;\label{ALEQ}\\ 
b_{i,k}(b,b') \,\equiv\,b_{i,k}\Lb p_i, \bar{p}_k, z_m\Rb \,\,&=&\, \h \frac{p_i- \bar{p}_k}{1 - p_i} -\frac{1}{2 z_m (1 - p_i)}\Lb (1- p_i) (1 - \bar{p}_k)  -   D_{i,k}\Rb;\label{BEEQ}\\
~~
D_{i,k} &=& \sqrt{4 p_i (1 - p_i) (1-  \bar{p}_k) z_m -  \Lb (1 - p_i) (1-  \bar{p}_k) - (p_i -  \bar{p}_k) z_m\Rb^2};\label{D}
\eea
  
  These equation are the explicit solutions to \eq{H03} and \eq{H0BC}.

\par{\it Amplitudes:}
In the following equations $ p_i \equiv p_i ( b')$ and $\bar{p}_k
 \equiv \bar{p_k}(\vec{b} - \vec{b} ')$.

~
$$z = e^{\Delta\,(1- p_{01})\,y}$$
~

$S_{ik} (a_{ik},b_{ik},z) \equiv S(a_{ik}(b,b'),b_{ik}(b,b'),z_m)$,  $ X_{i,k}( a, b,z)  \equiv  X(a_{ik}(b,b'),b_{ik}(b,b'),z_m)$

\begin{equation}
X(a_{ik},b_{ik},z)) =  \frac{a_{ik} + b_{ik} z}{1 + b_{ik} z}
\end{equation}

\begin{eqnarray}
&&SS_{ik}(a_{ik},b_{ik},z)=\\
&&-(a_{ik}-1) \text{Li}_2(-b_{ik} z)+a_{ik}
   \text{Li}_2\left(-\frac{b_{ik}
   z}{a_{ik}}\right)+(a_{ik}-1)
   \text{Li}_2\left(\frac{a_{ik}+b_{ik}
   z}{a_{ik}-1}\right)+\frac{1}{2} a_{ik} \log
   ^2((1-a_{ik}) b_{ik} z)\nn\\
   && -(a_{ik}-1) \log (b_{ik} z+1)
   \log ((1-a_{ik}) b_{ik} z)
  -\left(a_{ik} \log
   (z)-(a_{ik}-1) \log \left(-\frac{b_{ik}
   z+1}{a_{ik}-1}\right)\right) \log (a_{ik}+b_{ik}
   z)\nn\\
   &&  +a_{ik} \log (z) \log \left(\frac{b_{ik}
   z}{a_{ik}}+1\right)\nn
   \end{eqnarray}
   
   \begin{equation} \label{FIN}
   S_{ik}(a_{ik},b_{ik},z) \,\,=\,\,SS_{ik}(a_{ik},b_{ik},z) \,-\,SS_{ik}(a_{ik},b_{ik},z=1)   \end{equation}

~  
     The amplitude is  given by 

   \begin{eqnarray} \label{AIK}
\hspace{-1cm}&&A_{ik}(s, b)\,=\\
\hspace{-1cm}&&\,1  - \exp\Bigg( \frac{1}{p_{01}}\int \frac{m^2_1 d^2 b'}{4 \pi} \Big( S_{ik}(a_{ik},b_{ik},z_m) \,\,+\,\, a_{ik}(b,b') \Delta (1 - p_0) Y\Big)  - \int \frac{m_1^2 d^2 b'}{4 \pi}  \bar{p}_k( \vec{b} - \vec{b}',m_k)\,X(a_{ik},b_{ik},z_m) \Bigg)\nn
 \end{eqnarray}
 \subsection{ Physical observables.}
 
The physical observables in this model can be written as follows
\bea
\mbox{elastic~~ amplitude}:&~~&a_{el}(s,b )\,=\,i
 \Lb \alpha^4 A_{1,1}\,+\,2 \alpha^2\,\beta^2\,A_{1,2}\,
+\,\beta^4 A_{2,2}\Rb; \label{OBS1}\\
\mbox{elastic cross section}:&~~~&\sigma_{tot}\,=
 \,2 \int d^2 b\, 
a_{el}\Lb s, b\Rb;~~~\sigma_{el}\,=\, \int d^2 b \,|a_{el}\Lb s, b\Rb|^2;\nn\\
\mbox{elastic~~~\, slope}:&~~~&B_{el}\,=\, \frac{1}{2} \frac{\int b^2\, d^2 b \,\mbox{Im} A_{el}(Y, b)}{\int d^2 b \,\mbox{Im} A_{el}(Y,  b)};\label{OBBEL}\\
\mbox{optical~~~\, theorem}:&~~~&2 \,\mbox{Im} A_{el}(s,t=0)\,= \,2 \int d^2 b\, \mbox{Im} a_{el}(s,b)\,
=\,\sigma_{el} + \sigma_{in} \,=\,\sigma_{tot};\label{OBS11}\\
\mbox{elastic~~~\,cross section}:&~~~&\frac{d \,\sigma_{el}}{d t}\,\,=\,\,\pi\,|f(s,t)|^2;\,\,\,\,\,a_{el}(s,b)\,\,=\,\,\frac{1}{2 \,\pi}\int d^2 q\, e^{- i \vec{q}\cdot\vec{b}}\,f\Lb s,t\Rb \mbox{where}\,\, t\,=\,- q^2;\label{OBELT}\\
\mbox{single ~ diffraction}:&     & \sigma^{GW}_{sd}\,=\,2\,\int\,d^2 b \, \Lb\alpha \beta\{-\alpha^2A_{1,1}
+(\alpha^2-\beta^2)A_{1,2}+\beta^2 A_{2,2}\}\,\Rb^2;\label{OBS2}\\
\mbox{double \,\,diffraction}:&     &\sigma^{G W}_{dd}\,\,=\,\,\,\int d^2 b
 \,\, \alpha^4\beta^4
\left\{A_{1,1}\,-\,2\,A_{1,2}\,
+\, A_{2,2} \right\}^2.\label{OBS3}
\eea

It should be noted, that factor 2 in \eq{OBS2} takes into account the
  single diffractive dissociation of  the  two protons.

\section{ Comparison with   experimental data for proton-proton scattering}

\subsection{ The results of the fit in two channel model}
 
  As we have seen in the previous section, we introduce three dimensionless 
parameters: $\Delta$ - the intercept of the BFKL Pomeron, and $p_{01}$
 ($p_{02}$) - the amplitudes of the dipole-dipole scattering at low
 energies, and $\beta$ which  is related to the contribution of the 
diffractive
 production. 
  For $b$-dependence we suggested a specific form 
 (see \eq{IC}) which is characterized by the dimensional
 parameters: $m_i$. These  parameters  are determined
  by  fitting to the experimental data. We choose to describe
 five observables: total  and elastic cross sections, the elastic
 slope and single and double diffractions at low masses (see
 \eq{OBS1}-\eq{OBS3}).  
  
  The situation with the experimental data on the single and
 double diffraction production in proton-proton scattering at high 
energies,  is far from clear. It was  well
 summarized  in Ref.\cite{KMR}, to which we refer the reader.  We
 assume that the two channel model is able to describe  proton-proton 
 diffraction production in the entire kinematic region of
 produced mass. As  is shown in Ref.\cite{GUGU} for $\Delta\,>\,0$
 the integral over the  produced mass in diffraction  is convergent, and
 the Good-Walker mechanism\cite{GW} is able to describe the
 diffraction production both of small and large masses. However,
 the simple two channel model is a simplification, but we hope   to 
learn something by attempting to fit
   all available data  using  this simple model.

 From \fig{fit} one can see that we obtain quite a good description 
of  the data
 for $\sigma_{tot},\sigma_{el} $ and for the slope $B_{el}$  for
 $W \geq 0.5\,TeV$. Comparing with the one channel model of
 Ref.\cite{GLPPM} we start fitting from lower value
 of $W=0.5 \,TeV$ instead of $W=1\,TeV$.  We present The fitting 
parameters
  in Table I. One can see that both sets have the same
 qualitative features: the large value of the amplitude $A_{1,1}$
 and small values of other amplitudes. Note,  the values of
 parameters which describe this large amplitude turns out to
 be quite different in one  and two channels fits . Especially,
 this difference is seen in the value of $\Delta$ and masses
 ($m_1$ and $m_2$). The quality of the description of this
 model and the one channel model of Ref.\cite{GLPPM}  for
 $W = 0.5 \div 13 \,TeV$  are  more or less the same . However
 for $W> 1\,TeV$ the one channel model gives a better description.
 
 However, from \fig{sd}-a and \fig{dd}-a  it is clear that we
 failed to describe the data on the single and double diffraction
 production: roughly speaking we are able to describe  only  half of 
the
 values for single diffraction cross section.   
 Therefore,  the simple two channel model
 is not enough to describe the experimental data on the single
 diffraction production, in spite of the three new parameters
 that we have introduced. 
  Actually, we had the same situation in our   CGC 
motivated model of 
Ref.\cite{GLM2CH}. Hence, we can conclude that the fact that our model
 satisfies the unitarity constraints both in $t$ and $s$ channel
 unitarity is not sufficient, and  we need to search for  a more 
complicated
 model for the hadron structure.

  The values of parameters which led to the best agreement with the
 experimental data of  are shown in 
Table I. 
 The two  sets  of parameters  are quite different, but qualitatively
 they describe the data with large $A_{1,1}$ and small $A_{1,2}$ and $A_{2,2}$.

  Comparing these parameters with the resulting curves in \fig{fit} we
 see that shadowing corrections play an essential role. First,
  we  note that the value of  $\Delta_{\rm dressed}\,\,=\,\,\Delta
 \Lb 1 - p_{01}\Rb $ is rather large (about 0.5) in all variants. Recall,
 that  means that  $\Delta \approx 1$. Factor $(1 - p_{01})$ in
 $\Delta_{\rm dressed}$, stems from the enhanced diagrams that
 contribute to the Green function of the Pomeron. The resulting
 $\sigma_{\rm tot} \,\propto \,s^{\Delta_{\rm eff}}$ with 
 $\Delta_{\rm eff} \approx 0.07$.  The  reduction from
 $\Delta_{\rm dressed} $ to $\Delta_{\rm eff}$ occurs due to 
 strong shadowing corrections.

From \fig{dd}-a one can see that we failed to describe the
 double diffraction production.
 This reflects the situation which we had in our previous attempts
 to describe this process\cite{GLM2CH}.    The same problem  occurs
with  other groups (see, for example, Ref.\cite{KMR} and reference
 therein). The small size of the double diffraction cross section 
  in our model  occurs since the main contribution stems from
 the amplitude $A_{1,1}$ which is close to 1. Bearing this in mind we
 see that 
 $\sigma_{dd} \,\,\approx\,\,\frac{\beta^4}{\alpha^4} \sigma_{el}$ and
 since $\frac{\beta^4}{\alpha^4}$\,=\,(1/16 set I) (0.03 set II) the
 cross section turns out to be small.

\begin{table}[h]
\begin{tabular}{|l|l|l|l|l|l|l|}
\hline
Variant of the fit &$\Delta_{\rm dressed}$ & $p_{01}$ &$p_{02}$   & $m_1$ (GeV) &$m_2$(GeV)& $\beta^2$\\
\hline
I&0.488 $\pm$ 0.002&0.748 $\pm$ 0.002&0.005 $\pm$ 0.001&1.03 $\pm$ 012& 0.49 $\pm$ 0.08&0.134  $\pm$ 0.003 \\\hline
II& 0.499$\pm$ 0.01 &0.972 $\pm$ 0.02&0.166 $\pm$ 0.001& 1.05 $\pm$ 0.01 &1.44 $\pm$ 0.020& 0.2  $\pm$ 0.01  \\\hline
One channel &0.33 $\pm$ 0.03&0.489 $\pm$ 0.030 & &0.867  $\pm$ 0.005& & 0 \\\hline
\hline
\end{tabular}
\caption{Fitted parameters.$\Delta_{\rm dressed} = \Delta\Lb 1 - p_{01}\Rb$.}
\label{t2}
\end{table}

       \begin{figure}[ht]
    \centering
  \leavevmode
  \begin{tabular}{c c c}
      \includegraphics[width=6cm]{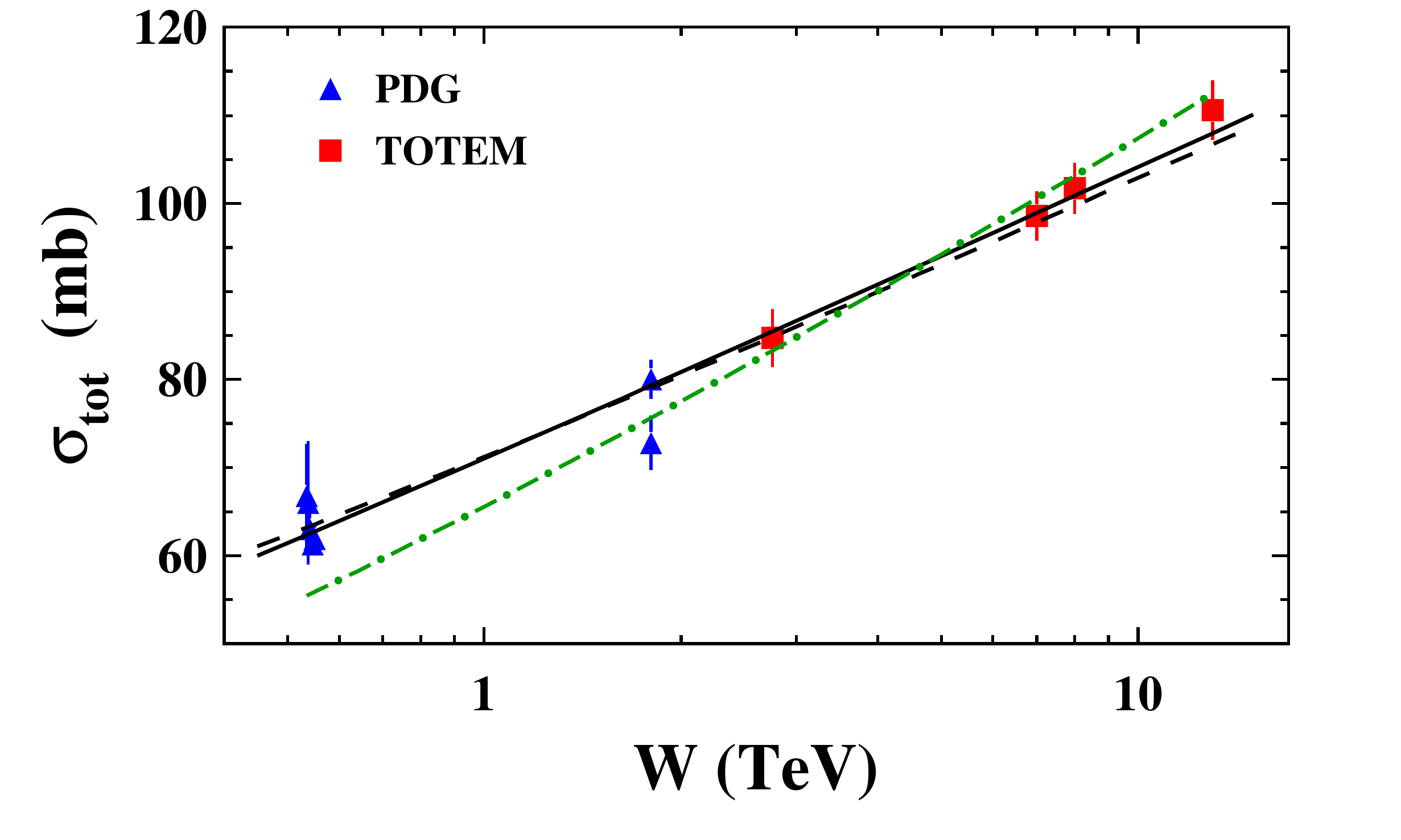}  &\includegraphics[width=6cm]{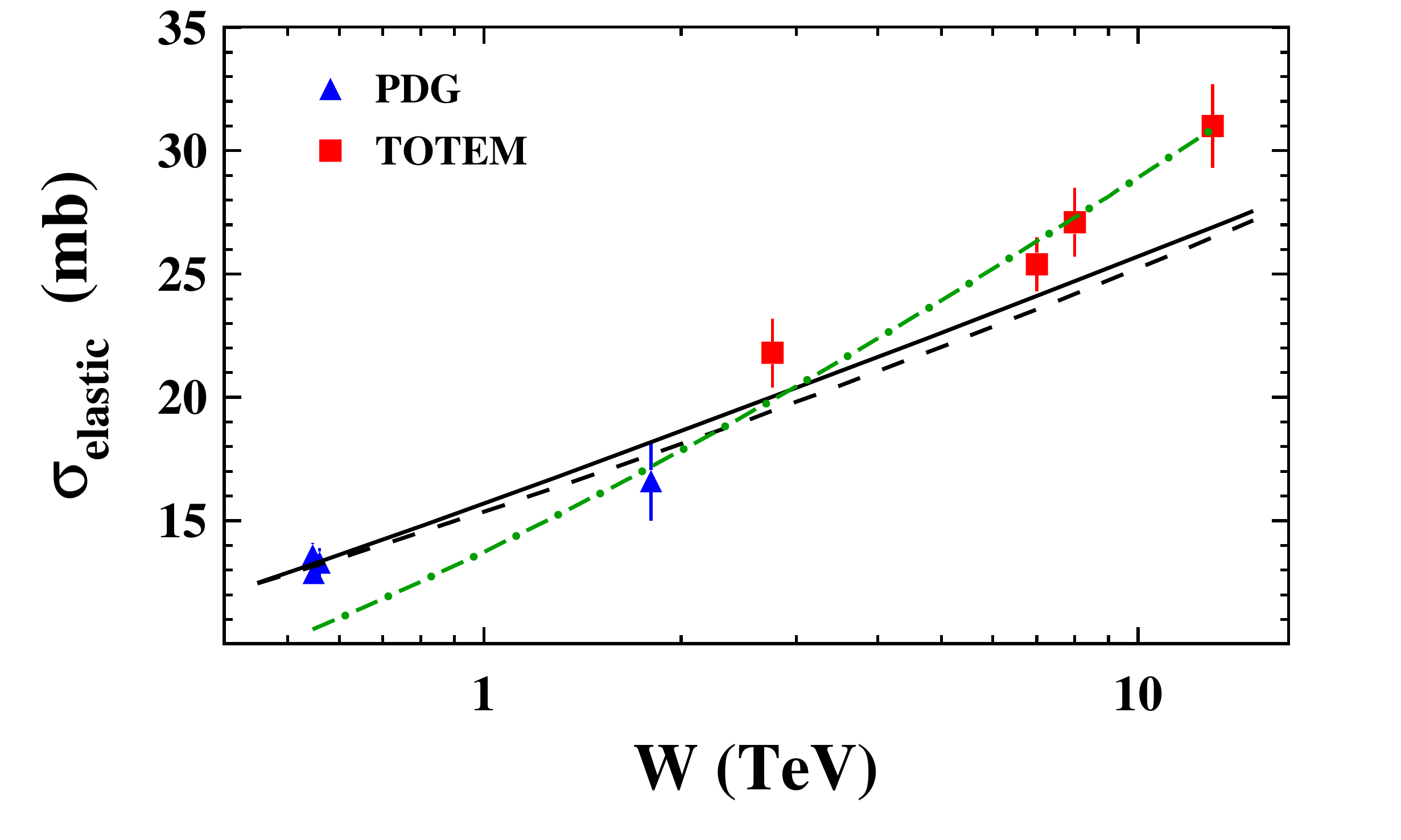}&\includegraphics[width=6cm]{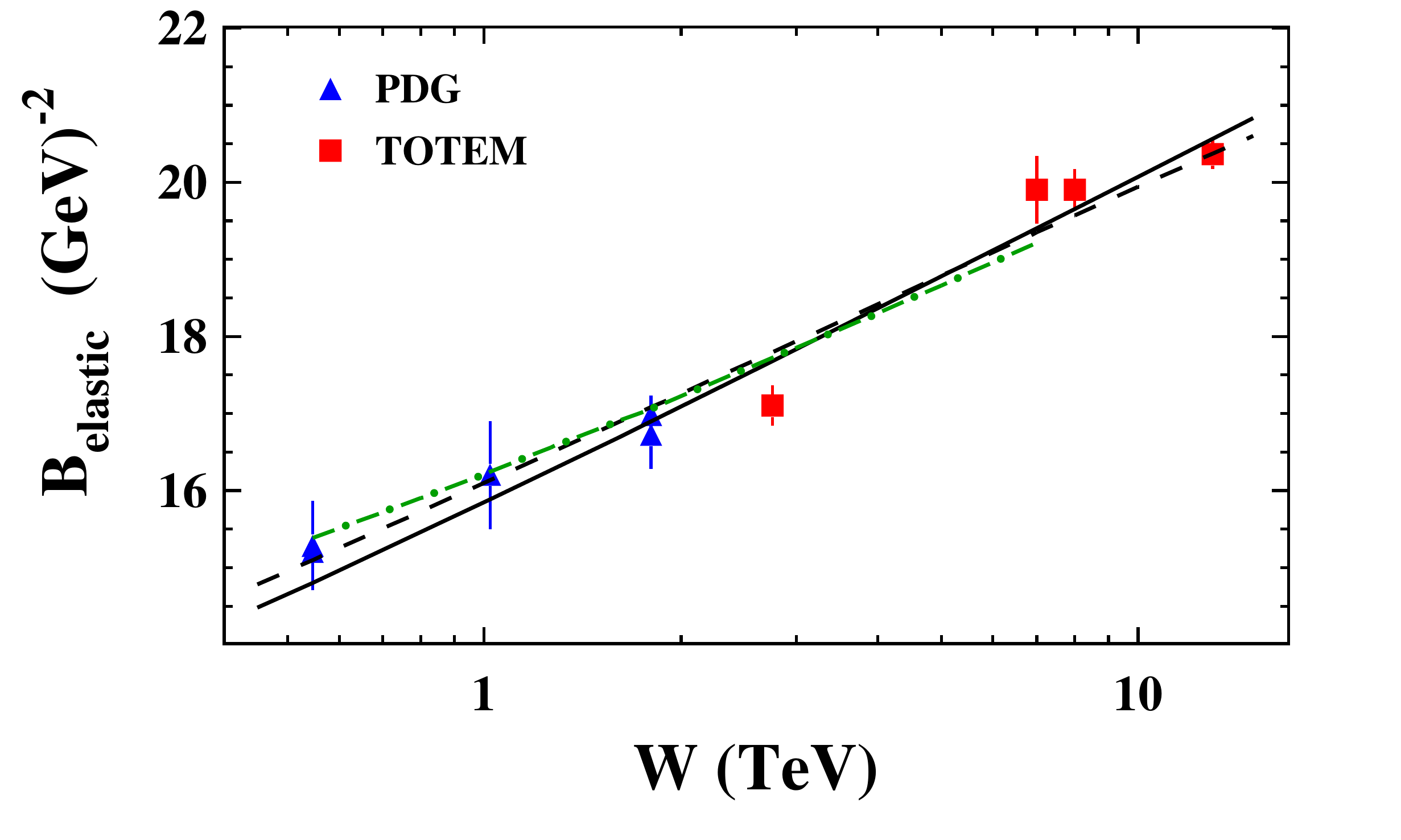} \\
      \fig{fit}-a&\fig{fit}-b& \fig{fit}-c\\
      \end{tabular}     
 \caption{ The energy behaviour of $\sigma_{tot},\sigma_{el}$ and the
 slope $B_{el}$ for proton-proton scattering in our model. 
The solid line describes the variant I in Table I while the
 dashed line  corresponds to variant II.
 Data are taken from Refs.\cite{PDG,TOTEMRHO}.   }
 \label{fit}
   \end{figure}



\subsection{ Diffraction production of large masses}
 
 The natural source of diffraction production, which we  neglected in
 our two channel model, is the production of large masses which can
 be reduced to the tripple Pomeron diagrams. \fig{dlm}-a and
 \fig{dlm}-b illustrate how the production of large mass is related
 to the exchange of the Pomerons. As we have discussed, our model is
 based on the theoretical approach that  describes the Pomerons and
 their interaction and, therefore,  we can estimate the contribution
 of the large mass to the diffraction production without  introducing
 any new parameters. In particular, the first diagram of \fig{dlm}-b
 takes the form:
 
 \beq \label{DLM1}
  \sigma^{\rm LM}_{\rm sd}\,\,=\,\,2\,\int^{Y}_0 \,d y'\,\int d^2 b'\, p_i \, G_\pom\Lb Y- y',\vec{b} - \vec{b}'\Rb \Gamma_{3\pom} \, G^2_\pom\Lb  y', \vec{b}'\Rb\, p^2_k
 \eeq
 where the triple  Pomeron vertex $\Gamma_{3 \pom}$ is known in our model,
 as well as the vertices of Pomeron interaction with the states 1 and 2
 ($p_1,p_2$).
$G_\pom\Lb y,b\Rb$ is the Green's function of the Pomeron. The factor 2 in
 front follows from the unitarity constraint for the Pomeron :
 $ \sigma_\pom \,=\,2\,G_\pom$.

       \begin{figure}[ht]
    \centering
  \leavevmode
      \includegraphics[width=11cm]{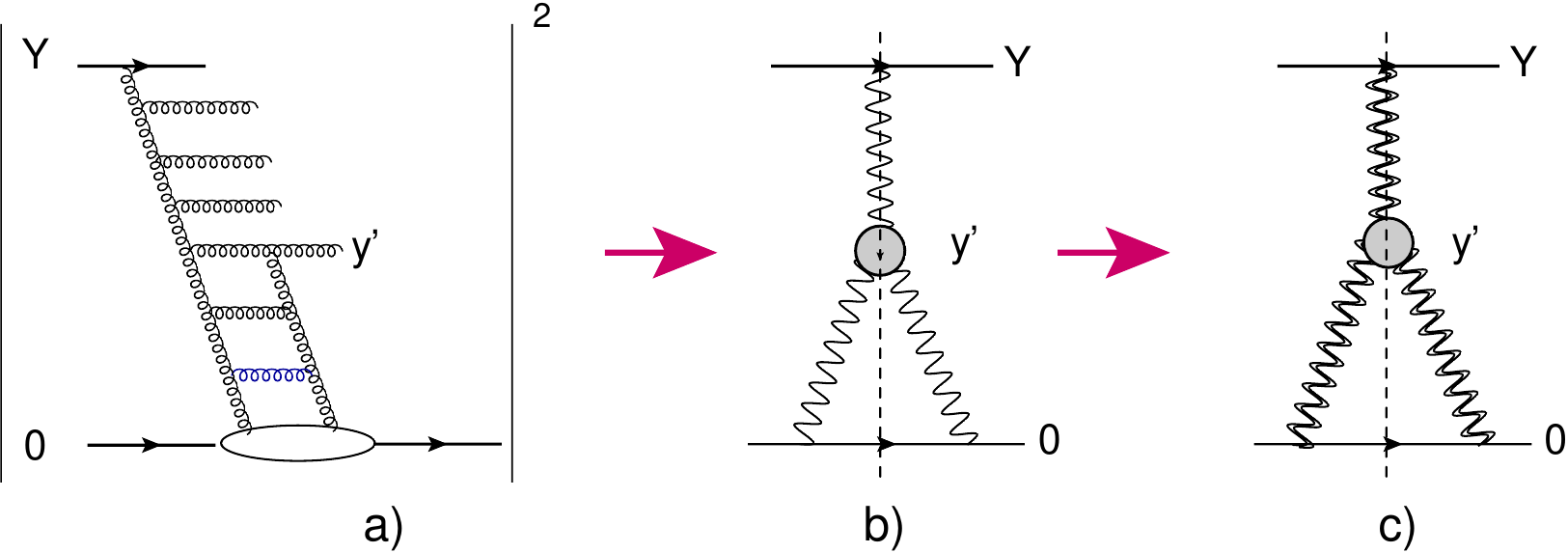}       
             \caption{The single diffraction production of large masses.
\fig{dlm}-a and \fig{dlm}-b present the first diagrams for the diffraction
 production. The blog represents  the triple Pomeron vertex.  
\fig{dlm}-c  
 corresponds  to the contributions of the `dressed ' Pomerons in our model.
 The wavy lines denote the Pomeron while the double wavy lines describe the
 resulting Green's function of the Pomeron in our model }      \label{dlm} 
   \end{figure}


We have the same expression as in \eq{DLM1} for the diagram of \fig{dlm}-c
 but we need to replace 
 the bare Pomeron Green's function by the resulting (`dressed') Pomeron
 Green's function ($G_\pom\Lb y,b\Rb\,\,\to\,\, G^{\rm dresssed}_\pom\Lb y,b\Rb $). In our model it turns out that easier to find not the resulting Green's function but the product $p_i\,G^{\rm dresssed}_\pom\Lb y,b\Rb $, which we will  denote 
 $\tilde{A}_{i, \pom}\Lb y, b\Rb$.
 We can find this amplitude from
 the general formulae of section II applying new initial conditions
 instead of \eq{IC}: viz.
 \beq \label{DLMIC}
 p_i\Lb b\Rb\,=\,p_{0i} S\Lb b,  m_i\Rb\,\,\,\mbox{for}\,\,\,\,i=1,2; ~~~~~~~
 p_\pom\Lb b\Rb\,=\,p_{0,\pom} \,S\Lb b, m_\pom\Rb
 \eeq
 
 From our model it follows that $p_{0,\pom}\,=\,p_{01}$, but the value
 of the mass $m_\pom$ should include the impact parameter dependence
 of the triple Pomeron vertex. It is  known that the radius of the
 triple Pomeron vertex is much smaller that the size of the proton. We
 choose the typical mass $m_\pom=3\,GeV$, which means that this radius
 is in about three times smaller than  the radius of the Pomeron-proton
 vertex.  We checked that the numerical estimates are not
 sensitive to the value of this mass.

       \begin{figure}[ht]
    \centering
  \leavevmode
      \includegraphics[width=8cm]{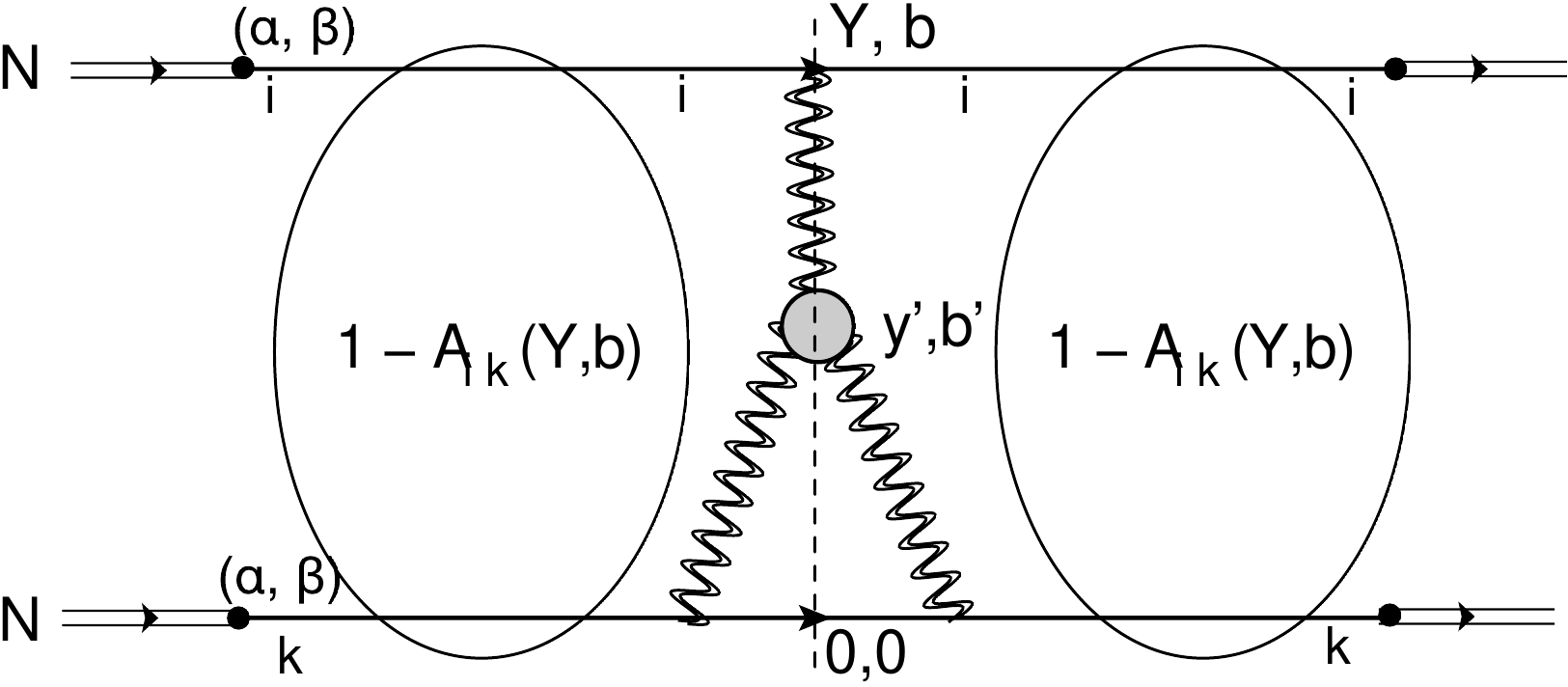}       
             \caption{The single diffraction production of large
 masses including the survival probability.
The wavy lines denote the Pomeron, while the double wavy lines
 describe the resulting Green's function of the Pomeron in our
 model . The black circles denote the transition
 $<\psi_h|\psi_i>\, =\,\alpha (i=1) $ or $\beta (i=2)$. }    \label{dlmgen} 
   \end{figure}

 
 However, we need to multiply \eq{DLM1} by the survival probability
 (see Refs. \cite{KMRREV,GLMREV} 
 for a review). Indeed, together with the processes shown in \fig{dlm} a
 number of parton showers can be produced    and gluons (quarks) from
 these showers will produce
 additional   hadrons  which, in particular, can fill the rapidity
 gap ($y'$ in \fig{dlm}). The survival probabilty factor $\sp$  gives
 the fraction of the processes in which the production of the parton
 showers are suppressed.
 Finally, the contribution of the single diffraction is given by the
 following expression (see \fig{dlmgen}):

 \beq \label{DLM2}
  \sigma^{\rm LM}_{\rm sd}\,\,=\,\,2\,\int^{Y}_0 \,d y'\,\int d^2 b'\,
  \, \tilde{A}_{i,\pom}\Lb Y- y',\vec{b} - \vec{b}'\Rb  
\tilde{A}^2_{k,\
\pom}\Lb  y', \vec{b}'\Rb\, \Big( 1\,\,-\,\,A_{i, k}\Lb Y,b\Rb\Big)^2
 \eeq

       \begin{figure}[ht]
    \centering
  \leavevmode
      \includegraphics[width=8cm]{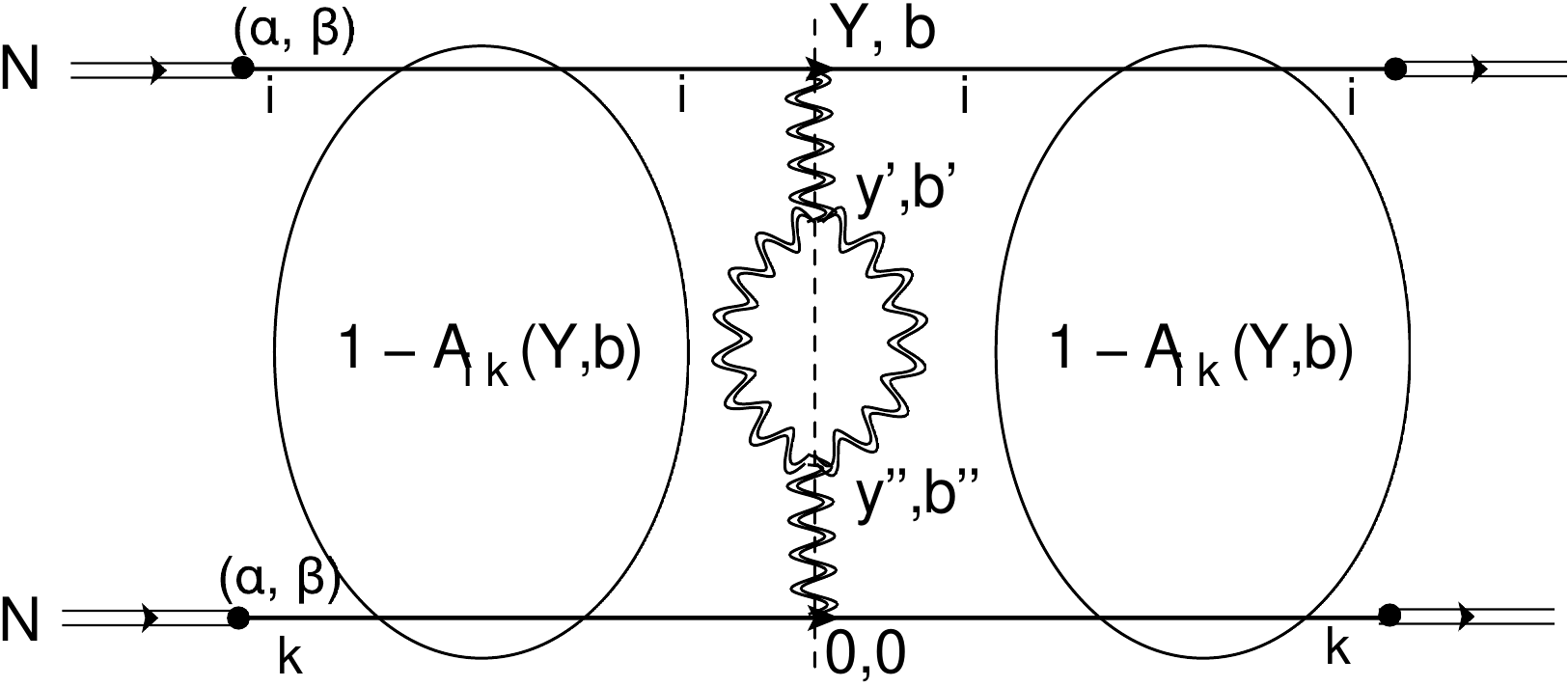}       
             \caption{The double diffraction production of large masses
 including the survival probability.
The double wavy lines describe the resulting Green's function of the
 Pomeron in our model. The black circles denote the transition
 $<\psi_h|\psi_i>\, =\,\alpha (i=1) $ or $\beta (i=2)$. }      \label{ddgen} 
   \end{figure}

The main contribution to $  \sigma^{\rm LM}_{\rm sd}$ for set I, stems
 from $A_{1,1}$, in spite of small values of $\sp$, since all other
 amplitude are small. For set II $A_{1,2}$ leads to the largest cross
 section due to large $\sp \approx 0.8$.

 For double diffraction in the region of large masses we can write the
 following formula which follows directly from \fig{ddgen}:
 
  \bea \label{DLM3}
&&  \sigma_{\rm dd}\,\,=\\
&&4\,\int^{Y}_0 \,d y'\,\int^{y'}_0 d \,y''\,\int d^2 b'\, \int d^2 b''\, \, \tilde{A}_{i,\pom}\Lb Y- y',\vec{b} - \vec{b}'\Rb  \tilde{A}^2_{\pom,\pom}\Lb  y' - y'', \vec{b}' - \vec{b}''\Rb\,\tilde{A}_{k,\pom}\Lb y'',\vec{b} - \vec{b}'\Rb \Big( 1\,\,-\,\,A_{i, k}\Lb Y,b\Rb\Big)^2\nn
 \eea


\subsection{Diffraction production: comparison with the experimental data}
 
 
 \fig{sd}-a shows a comparison of our results compared to  the single 
diffraction
 production data, taken from Ref.\cite{KAS}  and which are
  shown in \fig{sd}-b. One can see that the description is not very good
 at $W\,\approx\,0.5\,TeV$.   The reason for this is that, the integration
 over $y'$ in \eq{DLM2}  leads to the amplitude
 $A_{i,\pom}$ and $A_{k,\pom}$ enter at energies smaller than $W=0.5 
\,TeV$. 
 We cannot describe these energies in our model. For larger energies the
 phase space that corresponds to the unknown region of energies gives  
much smaller
  contributions.

 The TOTEM value of the single diffraction cross section is $9.1\pm 2.9 $
 (see Ref.\cite{KAS}) , while our estimates lead
 to $\sigma^{\rm smd}_{\rm sd} = 12-13 mb$.  As  can  be  seen from 
\fig{sd}-a and \fig{sd}-b,  our model leads to  values of the
 single diffraction cross section, which are  close to our predictions
from the  CGC motivated model of Ref.\cite{GLM2CH} (the curve GLM 
in 
\fig{sd}-b). We refer the reader to Ref.\cite{KMR} in which the
 situation with tensions between different  experimental  groups on 
the single
 diffraction cross section, has been discussed.
 
       \begin{figure}[ht]
    \centering
  \leavevmode
  \begin{tabular}{c c }
      \includegraphics[width=8cm]{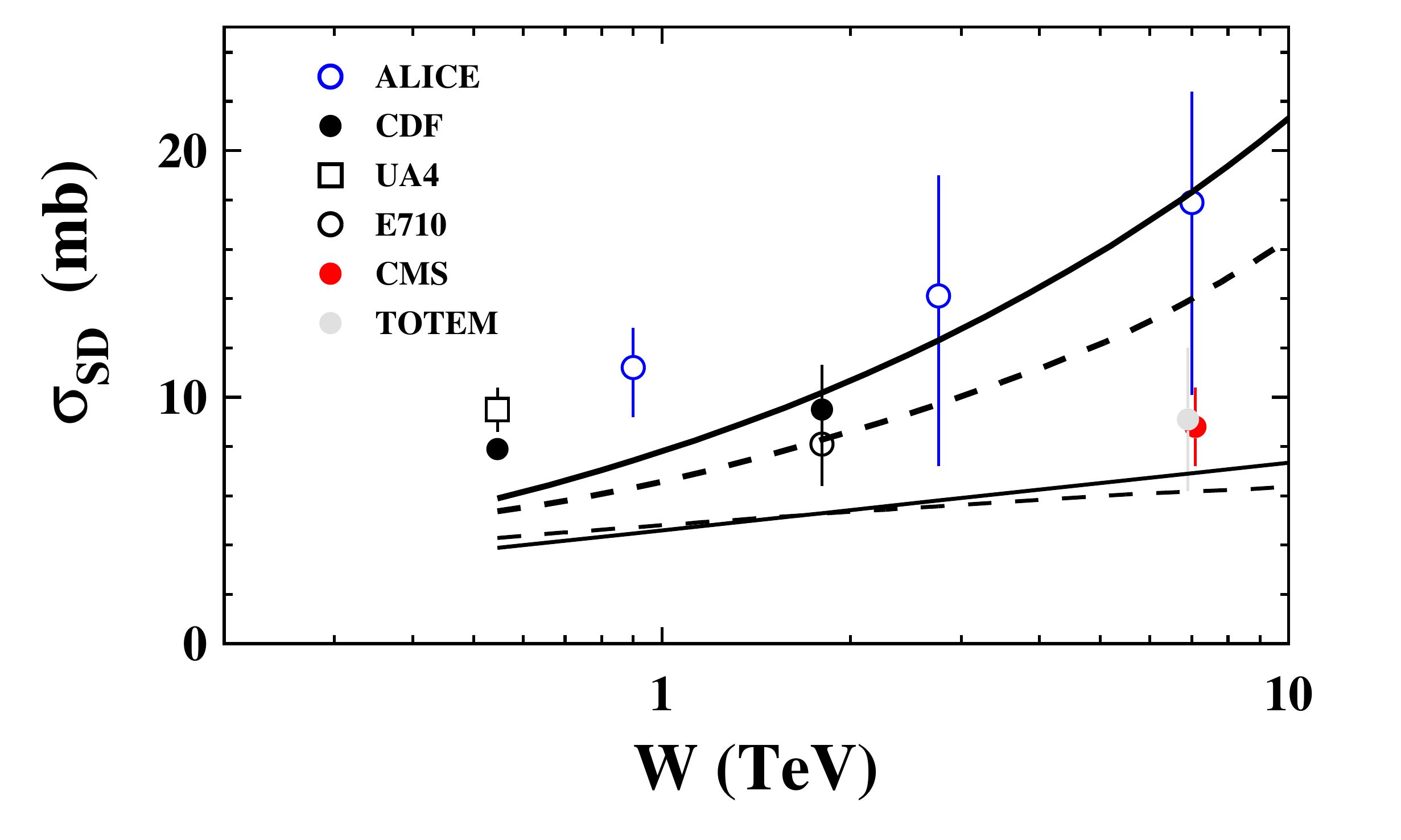}  &\includegraphics[width=9cm]{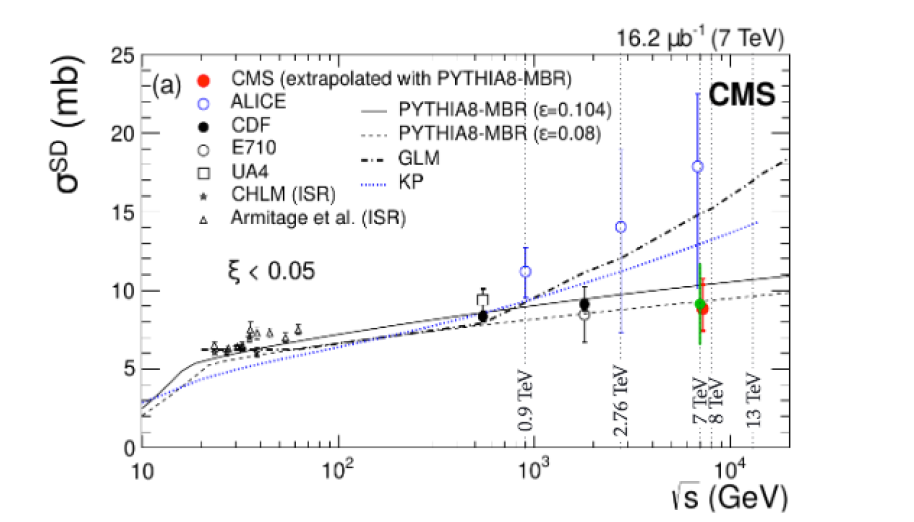} \\
      \fig{sd}-a&\fig{sd}-b\\
      \end{tabular}    
 \caption{  The single diffraction cross section as function of energy
 $W = \sqrt{s}$:
 our description of the data with $W\geq\,0.5\,TeV$ (\fig{sd}-a) and 
the experimental data from  Ref.\cite{KAS} (\fig{sd}-b). The  solid 
 and dashed lines in \fig{sd}-a  correspond to set I and set II (see
 Table I),respectively. The upper lines show the sum of small and
 large masses diffraction, while the lower ones present the small
 mass diffraction that were fitted in our model.
The data of
 all experimental groups were extrapolated to the region $M^2
 \leq 0.05\,s$ using the Pythia  Monte-Carlo programs as  is shown
 in \fig{sd}-b. $M$ is the mass of  hadron produced in   single
 diffraction. The data  in \fig{sd}-a are taken from Ref.\cite{KAS}
 and we refer to this paper (especially to Ref.[9] in it).  The curves
 in \fig{sd}-b marked as GLM are taken from Ref.\cite{GLM2CH} and that as 
KP is
 from Ref.\cite{KAPO}. }
 \label{sd}
   \end{figure}


       \begin{figure}[ht]
    \centering
  \leavevmode
    \begin{tabular}{c c }
   \includegraphics[width=8cm]{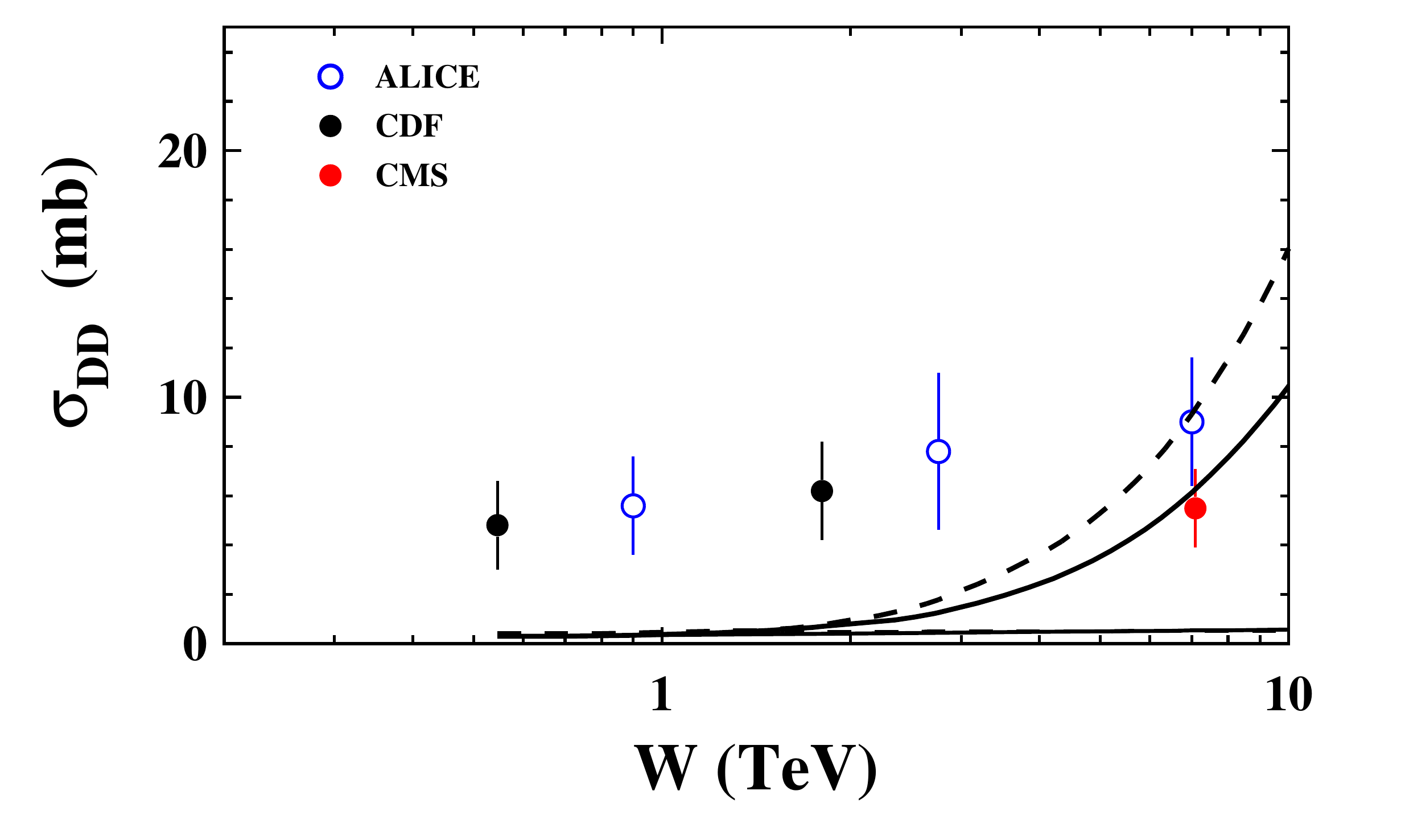}  &\includegraphics[width=8.5cm]{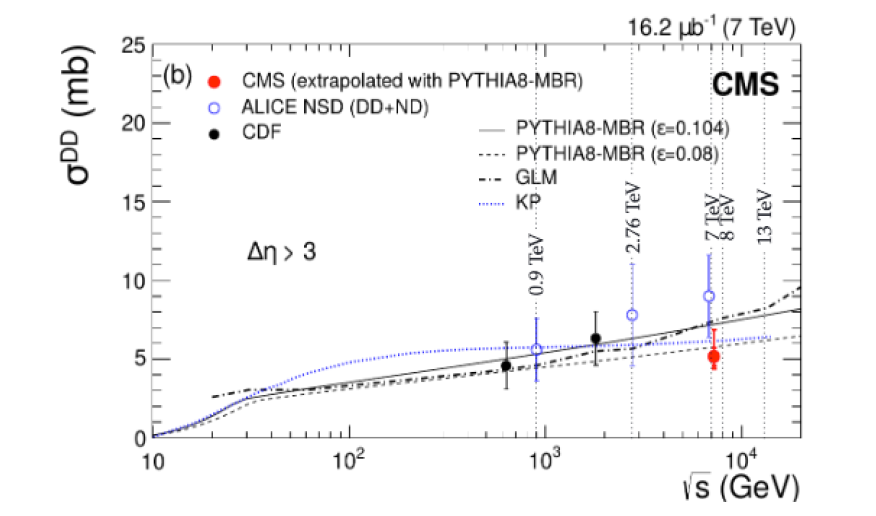} \\
      \fig{dd}-a&\fig{dd}-b\\
      \end{tabular}     \caption{   The double diffraction cross
 section as function of energy
 $W = \sqrt{s}$.The  solid  and dashed lines in \fig{dd}-a correspond to
 set I and set II (see Table I),respectively. The upper lines show the
 sum of small and large masses diffraction, while the lower ones present
 the small mass diffraction that were fitted in our model.  The experimental
 points, were taken from Ref.\cite{KAS}.   }
 \label{dd}
   \end{figure}

 The description of the double diffraction is very poor.
 The large mass diffraction leads to a large double diffraction cross
 section at high energies, and we  cannot reproduce the values
 of $\sigma_{dd}$ at lower energies. 
 
 Concluding this section we can claim that the large mass diffraction
 leads to a considerable contribution, which in this model increases
 rapidly with energy. This is a direct consequence of the large
 value of  $\Delta_{\rm dressed}$ in our model (see Table I).
 For $W \,>\, 0.5 \,TeV$, this increase is damped by large
 shadowing corrections, but in the formulae for diffraction production
 of large masses,  includes  energies which are less than 
$W\,=\,0,5$,
 where the strength of shadowing corrections  is not sufficient
  to lead to a reasonable effective $\Delta$.

\begin{boldmath}
 \section{Dependence on impact parameters}
\end{boldmath}

In \fig{ael} we plot the  scattering amplitudes as a function of the 
 impact parameter b.
One can see that the two channel model generates a very interesting and
 unexpected structure. One amplitude $A_{11}(b) $   has reached the
 unitary limit  $A_{11}\Lb b=0\Rb = 1$  at $W=0.5 \,TeV$ and shows
 the increasing  of the radius of the interaction, with energy.
The two other amplitudes are far  from the unitarity limit even at
 ultra high energy $W =  100\,TeV$. They increase as
 $W^{\Delta_{\rm eff}} $ with $\Delta_{\rm eff} \sim 0.1$.
The behaviour as a function of $b$ is also unexpected.
 Both $A_{11}$ and $A_{22}$ decrease monotonically    at large $b$, 
 while $A_{12}$ has a maximum which moves to larger values of
 $b$.   The value of the amplitude for this maximum increases
 as $W^{\Delta_{\rm eff}} $. On the other hand, $A_{12}\Lb b=0\Rb$
 is almost independent of $W$.

Such dependence of the amplitudes generate the elastic amplitude
 which is smaller than the unitarity limit even at very high
 energies (see \fig{ael}-a). This conclusion is in accord with 
the recent paper of Ref.\cite{GSS} in which it is demonstrated
 that  in the Miettinen-Pumplin \cite{MIPU}approach  the elastic
 amplitude $A_{el}\Lb b=0\Rb \approx 0.92 \,<\,1$ at $W = 57
 \,TeV$. Note, that this approach is ideologically close to ours and
 second, that in Ref.\cite{GSS} the entire set of  soft interaction
 data has been described  successfully.

In \fig{sd}-a we present the comparison between the elastic amplitude
 in our 2 channel model and in one channel model of Ref. \cite{GLPPM}.
 One can see that these two amplitudes have a  different behaviour
  both as a function of energy and  impact parameter. We believe that 
this
 figure demonstrates that the modeling of the non-perturbative  structure
 of the  hadron  is very important in understanding  high energy 
scattering.
  \fig{ssd}-b shows the behaviour of $ d \sigma_{\rm sd}/d b^2$ (see \eq{OBS2})
\beq \label{SD}
\frac{d \sigma_{\rm sd}}{d b^2}\,\,\,=\,\,\, \Lb\alpha \beta\{-\alpha^2A_{1,1}
+(\alpha^2-\beta^2)A_{1,2}+\beta^2 A_{2,2}\}\,\Rb^2
\eeq

One can see that this observable  decreases  very slowly  
with energy,
 and does not show a maximum at large $b$. Such  behaviour is quite
 different from what we obtain in CGC motivated model (see
 Ref.\cite{GLM2CH} Fig.7) and from the estimates of Ref.\cite{GSS}.

We believe that
the impact parameter and energy behaviours shown in \fig{ael}
 and in \fig{sd},  illustrate the fact that  the soft interaction
 at high energies could have a much richer structure than we previously 
 assumed.

       \begin{figure}[ht]
    \centering
    \begin{tabular}{c c c c}
  \leavevmode
      \includegraphics[width=4cm]{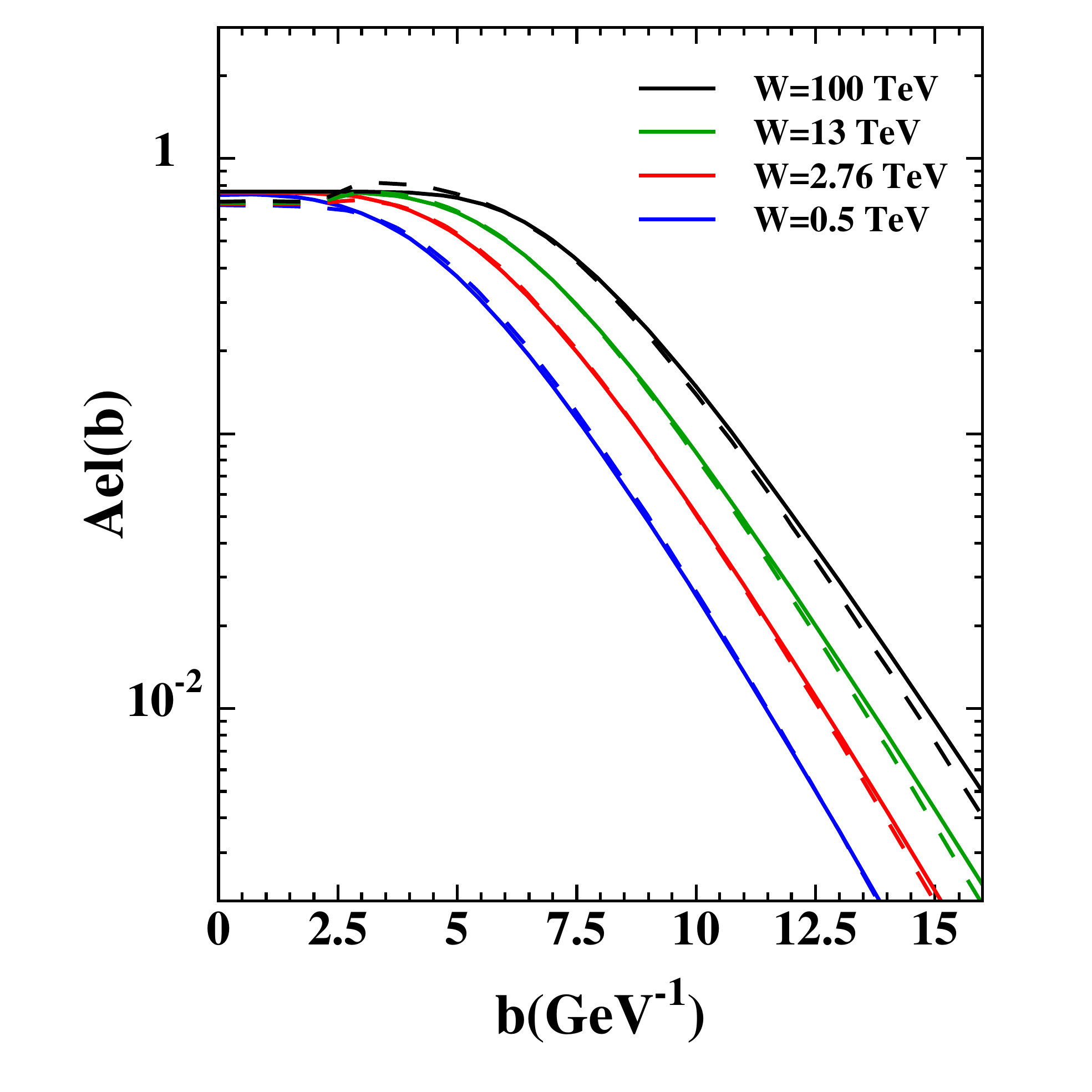} &    
  \includegraphics[width=4cm]{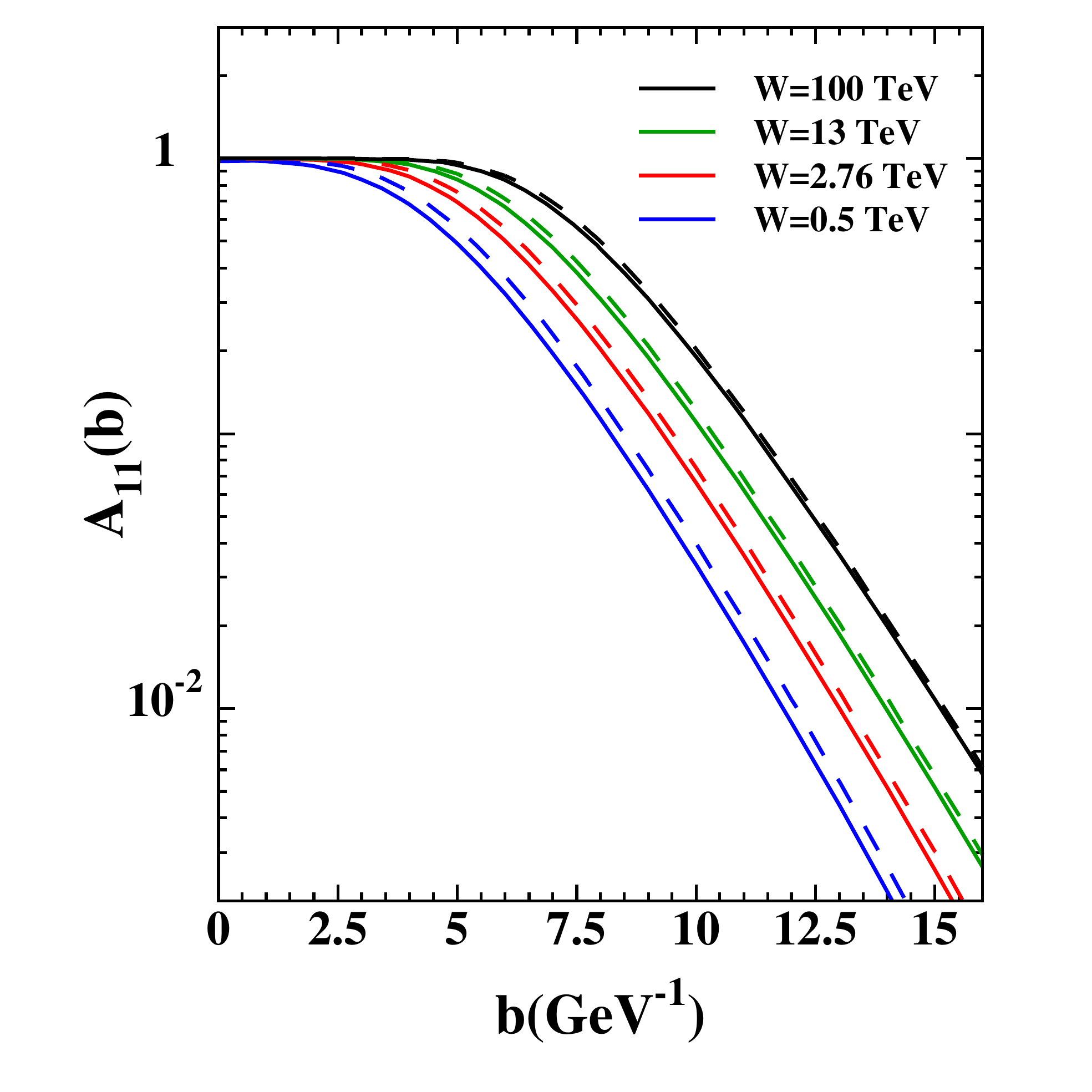} &
            \includegraphics[width=4cm]{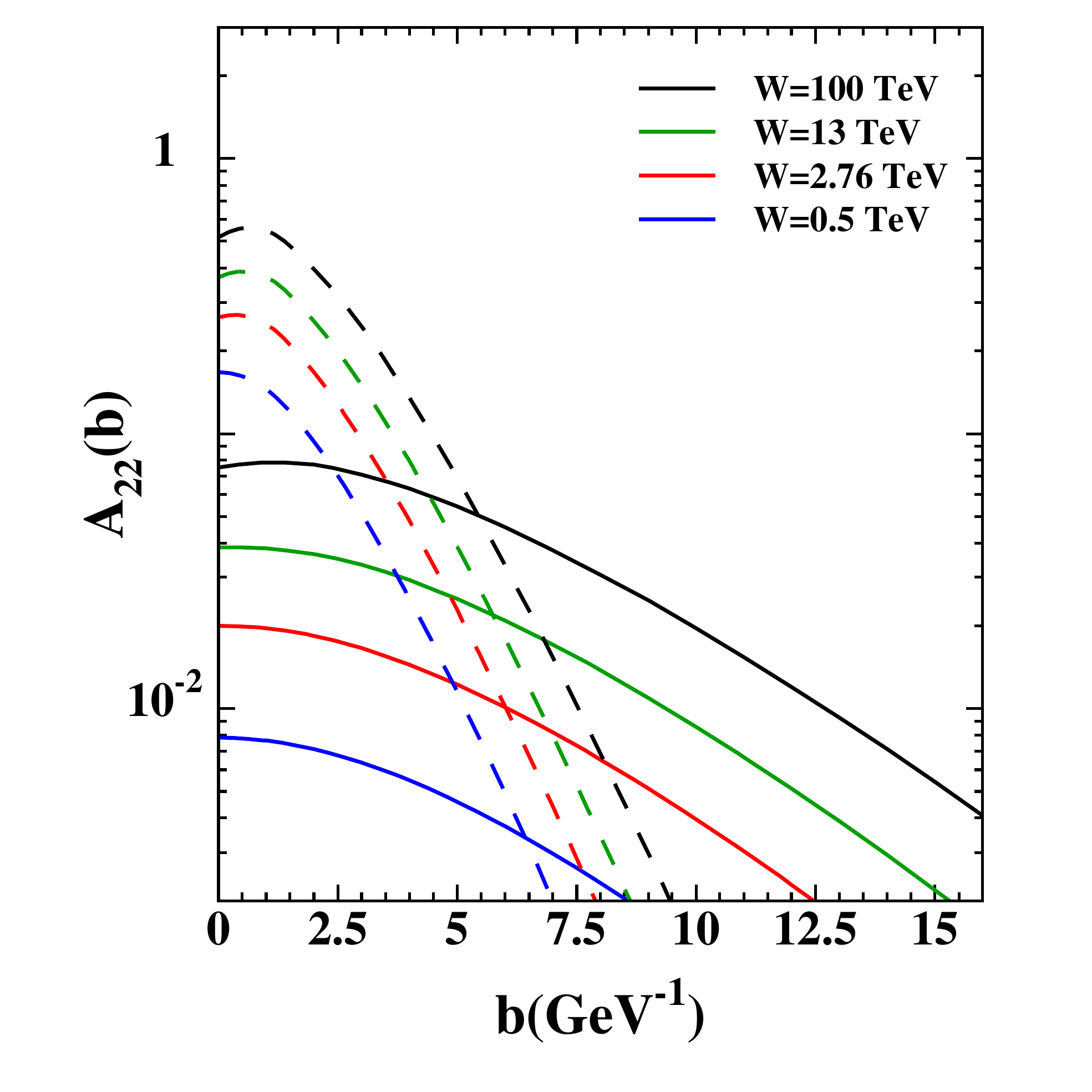} & 
    \includegraphics[width=4cm]{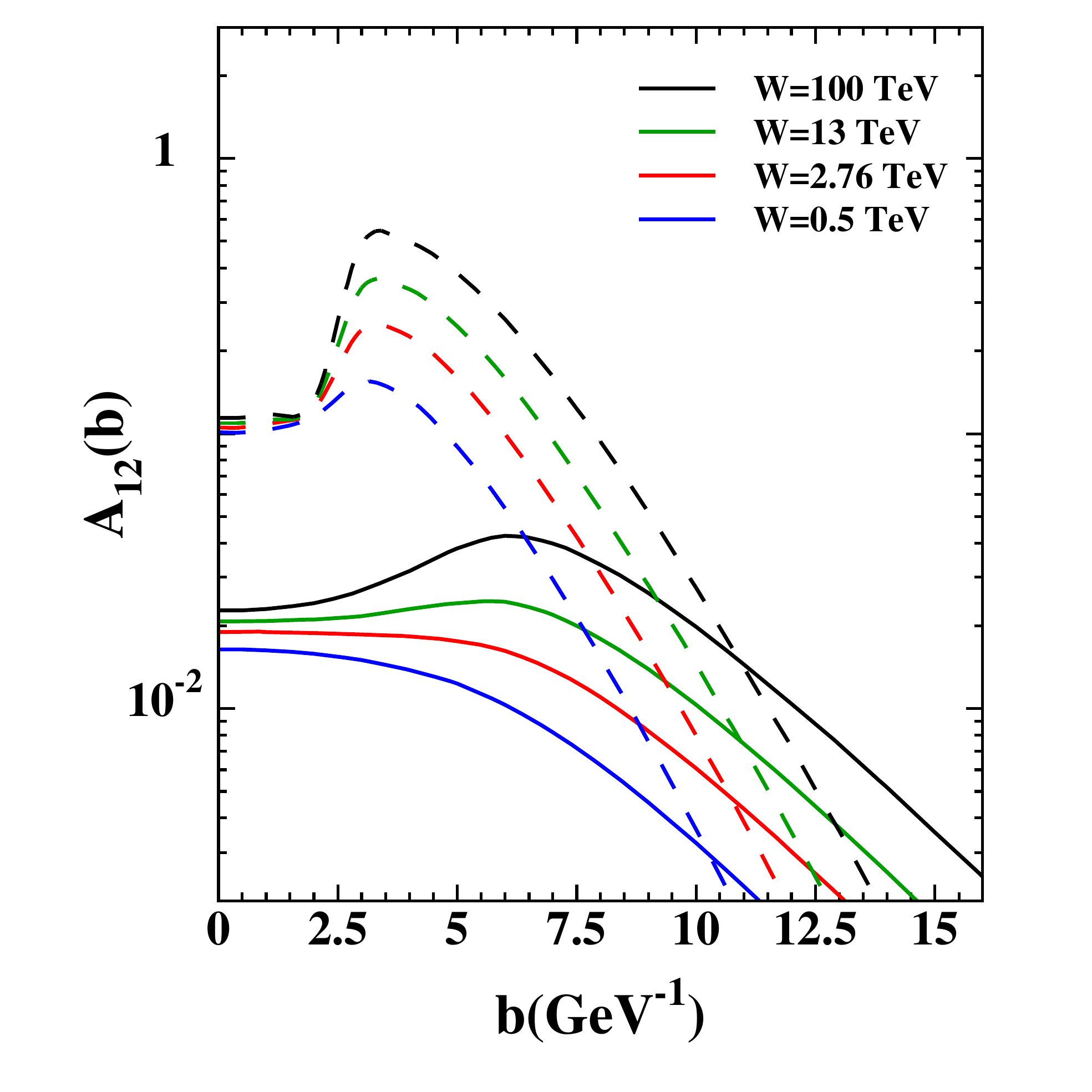}\\
  \fig{ael}-a & \fig{ael}-b & \fig{ael}-c&\fig{ael}-d\\  
  \end{tabular}        
             \caption{The  scattering amplitudes versus  impact 
parameter  $b$ for
 different energies:\fig{ael}-a $A_{el}$; \fig{ael}-b $A_{11}$,
\fig{ael}-c $A_{22}$,\fig{ael}-d  $A_{12}$. }      
\label{ael} 
   \end{figure}

~

~

       \begin{figure}[ht]
    \centering
    \begin{tabular}{c c c}
  \leavevmode
      \includegraphics[width=7cm]{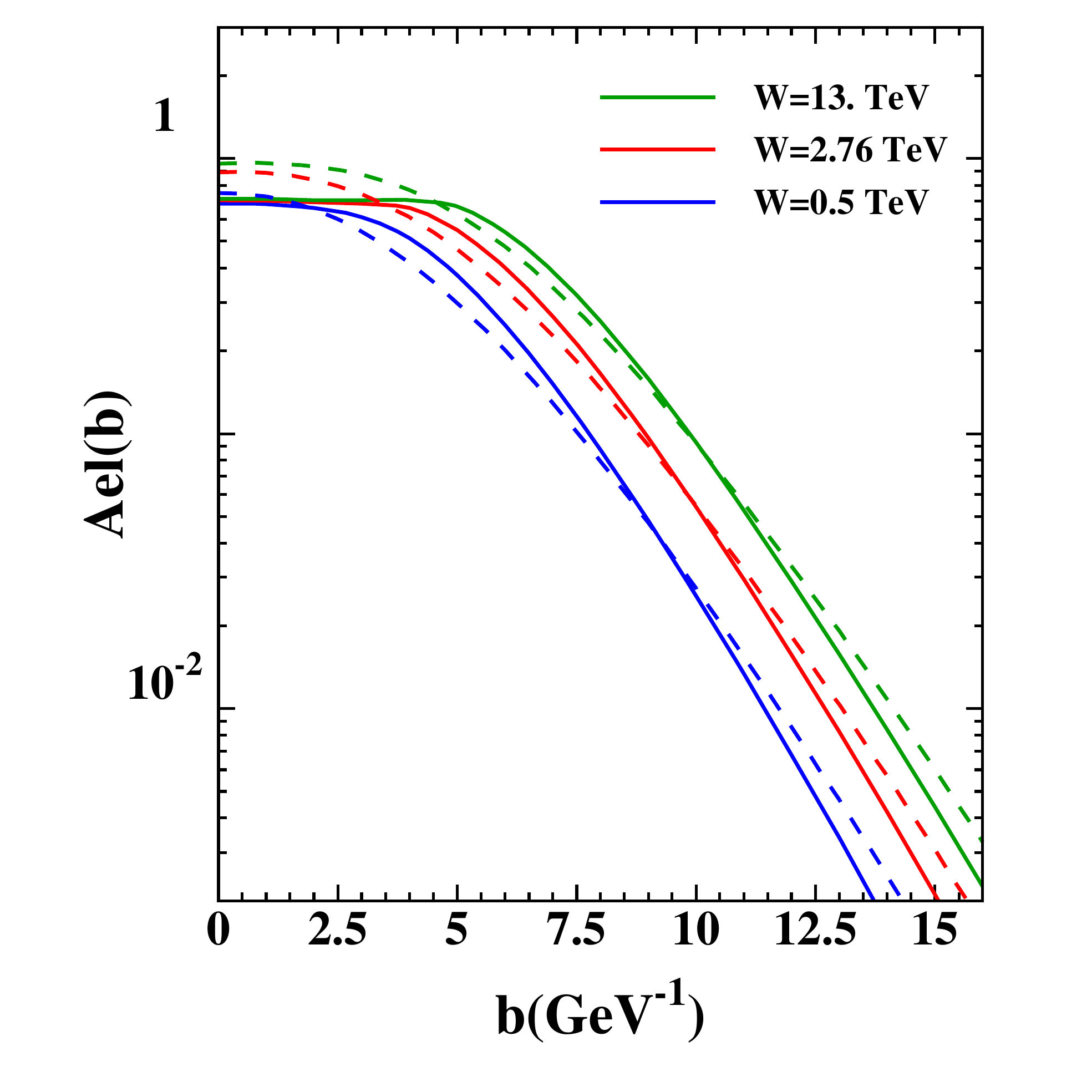} &   &   \includegraphics[width=7cm]{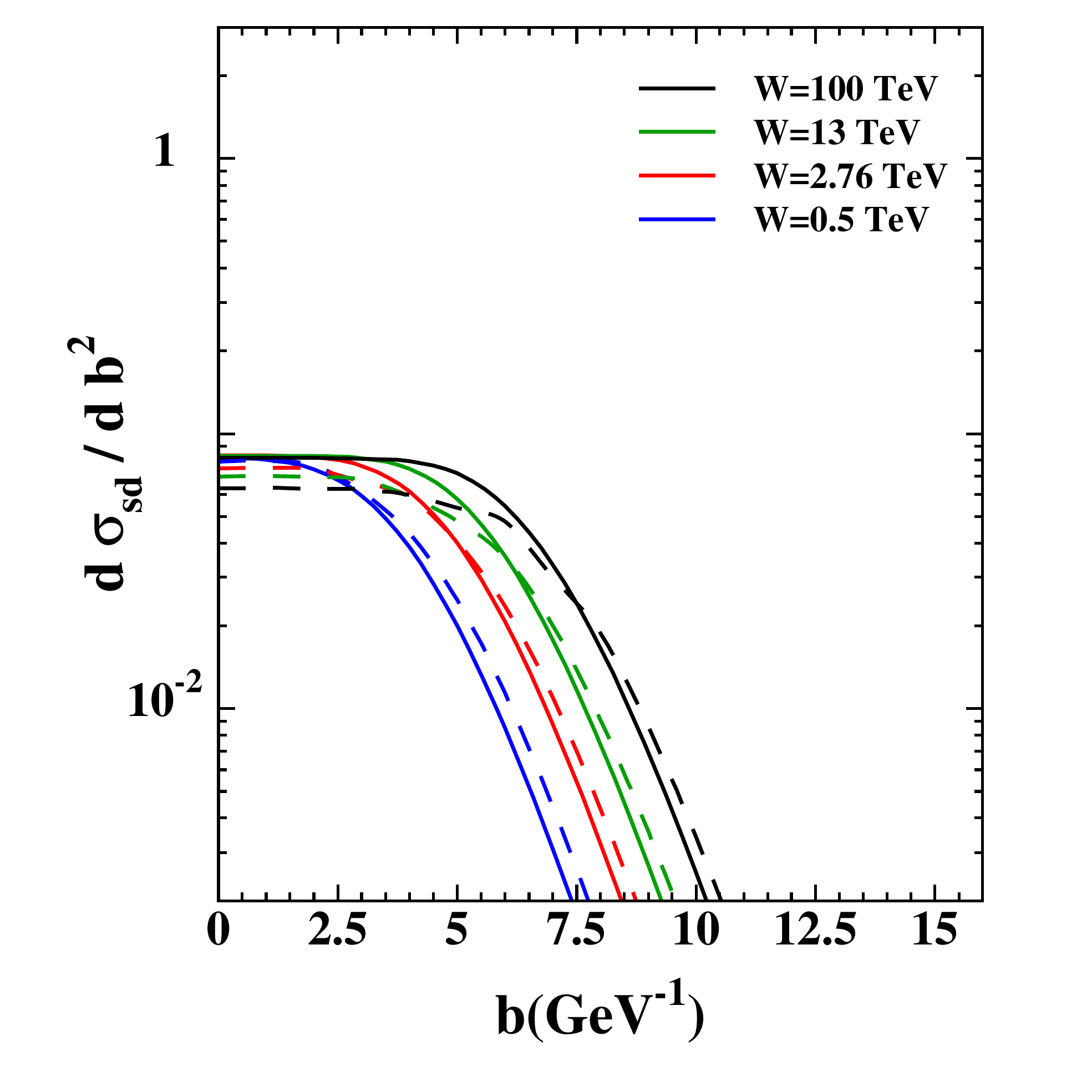} \\
  \fig{ssd}-a & ~~~~~&\fig{ssd}-b\\  
  \end{tabular}        
             \caption{The  scattering amplitudes versus
  impact parameter  $b$ for
 different energies:
             \fig{ssd}-a:   the elastic amplitudes for the
  one channel model of Ref.\cite{GLPPM}
 (dashed line) and for two channel model of this paper 
 (solid  line). For estimates in our model we used set I
  of parameters in Table I; \fig{ssd}-b 
 :     
$\frac{d
 \sigma_{\rm sd}}{d b^2}$ of \protect\eq{SD}  for the  variant two
 (solid line) and variant one (dashed line) set of parameters. }      
\label{ssd} 
   \end{figure}


\begin{boldmath}
 \section{Dependence of the elastic cross sections on $t$}
\end{boldmath}

 We  attempt  to describe the elastic cross section for $|t|=0 \div 
1\,GeV^2$ to 
check the rich structure present  in the impact parameter dependence, 
this stems 
from
 our model,  which  predicts the existence of a 
minimum
 in the elastic cross sections, however its position occurs at
 $|t |\,\approx\,0.3\,GeV^2$, which is much smaller than was
 observed experimentally by TOTEM collaboration\cite{TOTEMLT}. 
 
 Assuming that this discrepancy  is due to the simplified 
form of
 $b$ dependence of our amplitude which is given by \eq{IC}, we
 changed the initial conditions of \eq{IC} to the following equations
 \begin{equation} \label{IC1}
p_i(b') = p_{0 i} \,S(b',m_i,\mu_i,\kappa_i)~~~\mbox{with}
~~S(b,m_i,\mu_i,\kappa_i)=  \Lb 1 - \kappa_i\Rb \Lb m_i \,b\Rb^{\nu_1} K_{\nu_1}(m_i \,b )\,\,+\,\,\kappa_i \frac{\Lb m_i \,b\Rb^{\nu_2} K_{\nu_2}(\mu_i \,b )}{2^{\nu_2\,-\,1} \,\Gamma\Lb \nu_2\Rb}
\end{equation}

As we have seen in \fig{fit} our two channel model does not give a 
good description of 
  the elastic cross section. Bearing this in mind  we made a fit  
using
 the   one channel model in which 
$p_{01}\neq 0$
 but  $p_{o2}=0$. In the Table II we present the parameters that we  found
 for the fit. \fig{dsdt} shows the comparison with  TOTEM data
 of Ref.\cite{TOTEMLT}. One can see that we  obtain    good agreement
 with the experimental data for $|t|\,<\,|t|_{min}$ and for
 $|t|\,>\, |t|_{min}$. However, for $ |t| \approx\,|t|_{min}$ the real 
part 
of the scattering amplitude  turns out to be small, and we obtain
a value of the $d \sigma_{el}/d \,t$ approximately   an  order of 
 magnitude smaller than the experimental one.  It should be stressed that
 we do not use  any of  the simplified approaches to estimate the 
real part of
 the amplitude, but using our general expression of \eq{AIK}  for $A_{i 
k}\Lb s,t\Rb$, we consider the sum $A_{ik}\Lb s,+ i \epsilon t\Rb +
 A_{ik}\Lb u-i \epsilon,t\Rb$, which corresponds to positive signature,
 and calculated the real part of this sum.

 In \fig{dsdt} we estimate the contribution of the $\omega$ -reggeon , 
using the 
 description  taken from the paper of Ref.\cite{DOLA}(note the difference
 between green dashed line and the blue solid curve). This
 contribution is  small, and can be neglected.
       \begin{figure}[ht]
    \centering
  \leavevmode
      \includegraphics[width=11cm]{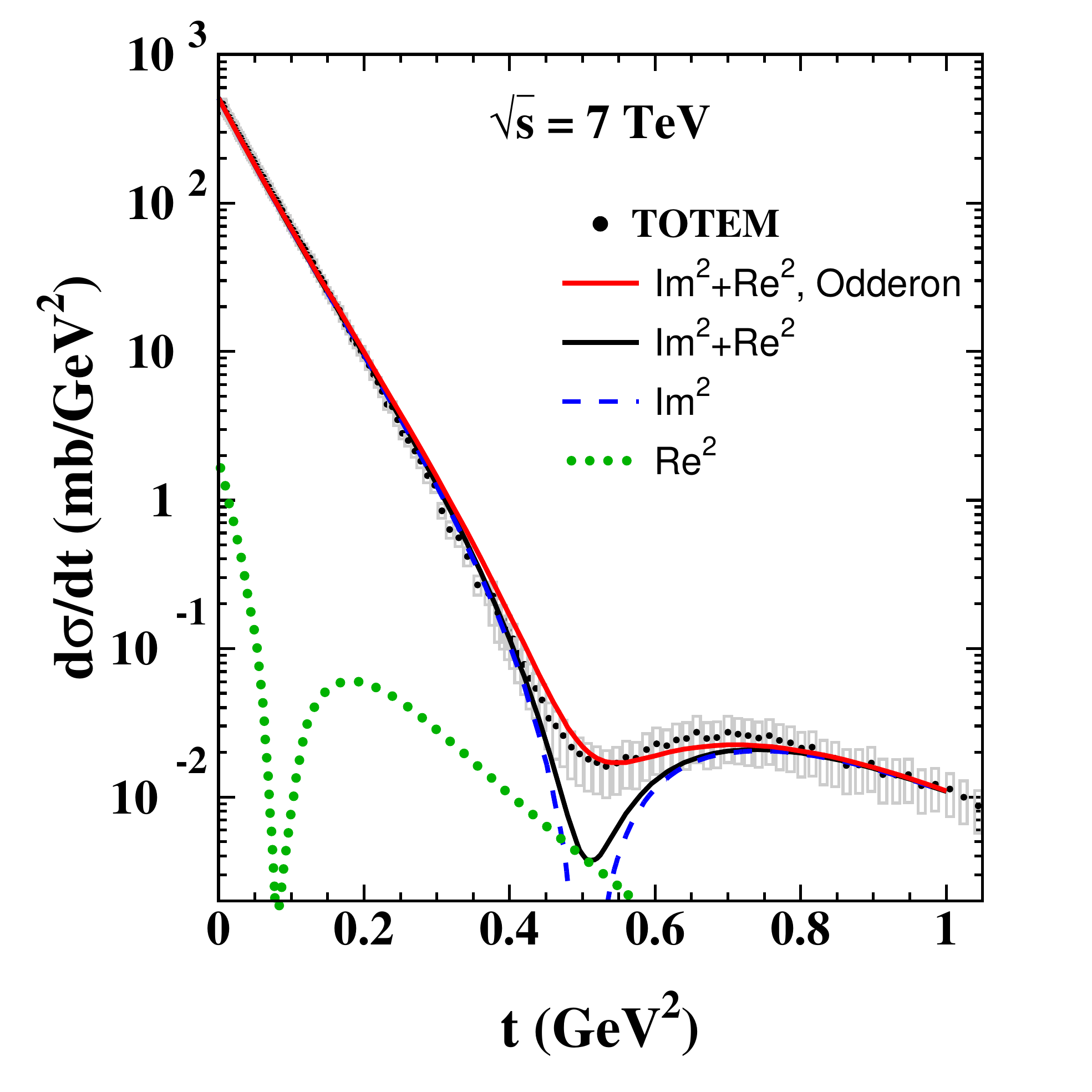}       
             \caption{$d \sigma_{el}/dt$ versus $t$. The black green line
 describes the result of our fit. The dashed  line corresponds to the
 contribution of the imaginary part of the scattering amplitude  to the
 elastic cross section. The dotted line relates to the real part of our
 amplitude.The red solid line takes into account  the contribution of the
 odderon to the real part of the $p p$ amplitude, as is shown in 
 \eq{ODD}.  The data , shown in grey, include systematic errors. They  are
 taken from Ref.\cite{TOTEMLT}. }      \label{dsdt} 
   \end{figure}


To evaluate the real part of the amplitude  we use the relation: However, 

\beq \label{DSDT1}
{\rm Re}A_{11}(s,t)\,\,=\,\,\h \,\pi\,\frac{\partial}{\partial\,\ln\Lb s/s_0\Rb}\, {\rm Im}A_{11}\Lb s,t\Rb|_{\eq{AIK}}
\eeq
\eq{DSDT1} correctly describes the real part of the amplitude only for
 small $\rho={\rm Re}A/{\rm Im} A$. In \fig{dsdt1} we plot
 the $d \sigma/dt$ with such estimates for the real part.
  The real part from    \eq{DSDT1} turns
 out to be  almost
 twice  larger than the experimental  data  in the vicinity of $t_{min}$.
  Therefore, at  the minimum, where 
 ${\rm Im}\, A \,\ll\,{\rm Re}A$, \eq{DSDT1} cannot  be used 
 for the real part. However, replacing \eq{DSDT1} by 
\beq \label{DSDT2}
{\rm Re}A_{11}(s,t)\,\,=\,\,\tan\Lb\rho\Rb\, {\rm Im}A_{11}\Lb s,t\Rb|_{ \eq{AIK}}
\eeq
we obtain the same  result, that the real part of the amplitude turns
 out to be too large. Actually,\eq{DSDT2} assumes that the scattering
 amplitude depends on energy as a power $A\Lb s,
 t\Rb\,\propto\,s^{2\,\rho/\pi}$. Our amplitude
 is a rather complex function of energy, and depends
 on $\ln(s)$.
 
       \begin{figure}[ht]
    \centering
  \leavevmode
      \includegraphics[width=10cm]{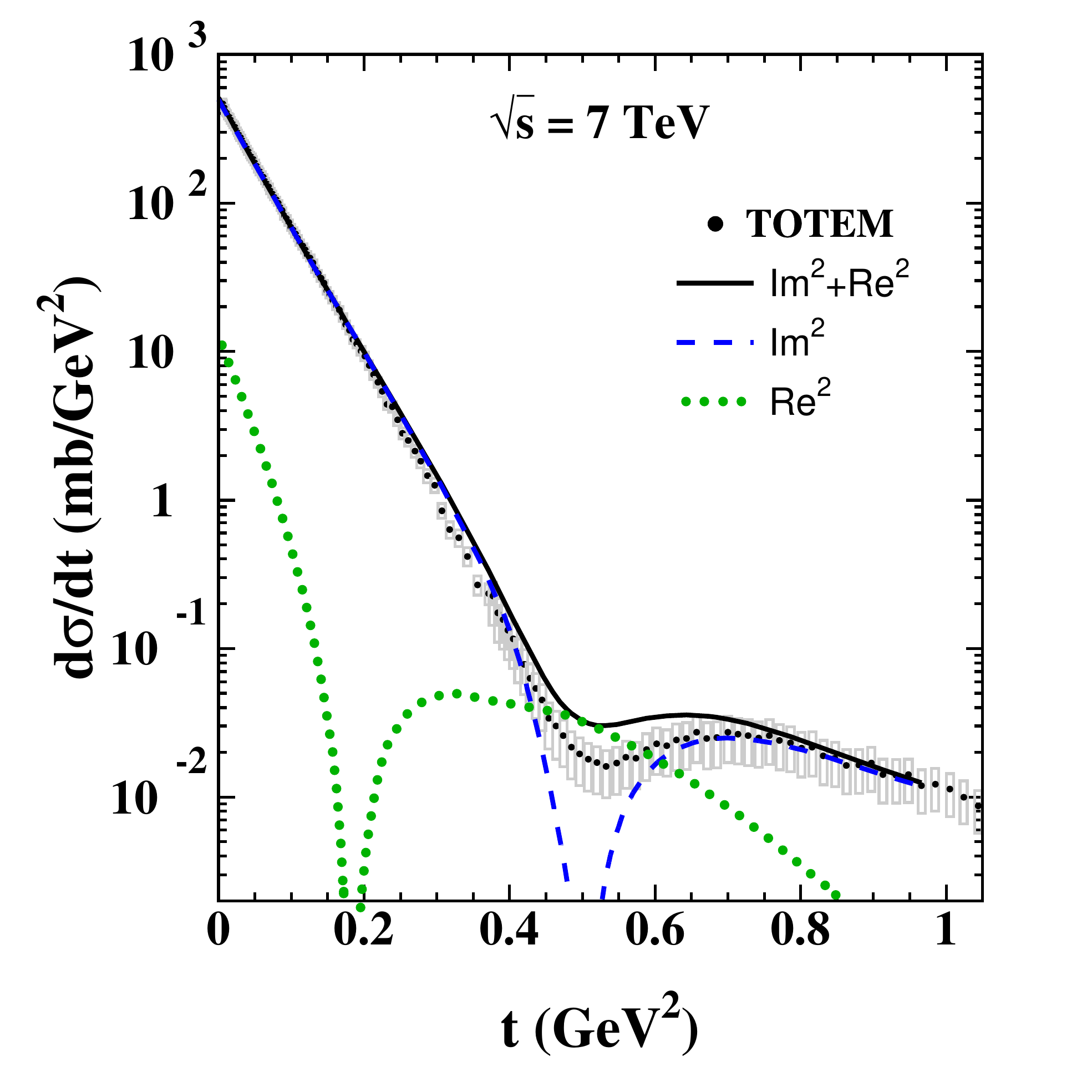}       
             \caption{$d \sigma_{el}/dt$ versus $t$. The solid  line
 describes the result of our fit. The dotted  line corresponds
 to the contribution of the real part of the scattering amplitude
  to the elastic cross section, which is calculated using \eq{DSDT1},
 with added contribution of the
 exchange of the  $\omega$ - reggeon,  which is taken from 
Ref.\cite{DOLA}. We do not show the contribution of the real part
 without the $\omega$-reggeon since it coincides with the dotted
 line. The dashed line is the contribution of the imaginary part
 of the amplitude.
 The data are taken from Ref.\cite{TOTEMLT})
              }      
\label{dsdt1} 
   \end{figure}


\begin{table}[h]
\begin{tabular}{|l|l|l|l|l|l|l|l|}
\hline
Variant of the fit &$\Delta_{\rm dressed}$ & $p_{01}$   & $m_1$
 (GeV) &$\mu_1$(GeV)& $\nu_1$& $\nu_2$&$\kappa_1$\\
\hline
one channel model  &0.48 $\pm$ 0.01&0.8 $\pm$ 0.05&0.860 &7.6344&0.9&0.1 &0.48\\\hline
\hline
\end{tabular}
\caption{Fitted parameters for $d \sigma_{el}/d t$ 
dependence.$\Delta_{\rm dressed} = \Delta\Lb 1 - p_{01}\Rb$.}
\label{t3}
\end{table}
 Concluding, we see that  to describe the TOTEM experimental data in 
the framework of our model, the contribution to
 the real part of the amplitude from the exchange of the odderon\cite{ODD}
  is needed. 
Hence, our
 estimates confirm the conclusions of Ref.\cite{ODDSC}. In
 \fig{dsdt} we plot the description of the elastic cross section in 
 which we  have added  the odderon contribution to the amplitude of 
\eq{AIK}
 (red solid curve in \fig{dsdt}):
 \beq \label{ODD}
 f\Lb s,t\Rb\,\,=\,\,f\Lb s,t; \eq{AIK}\Rb\,\,\pm\,\, \sigma_{\rm odd} e^{B_{\rm odd} \,t}
 \eeq
 where we consider a QCD odderon\cite{ODD}: the state with odd signature and
  with the intercept  $\alpha_{\rm odd}(t=0)=1$,  which contributes only
 to the real part of the scattering amplitude. The value of $\sigma_{\rm
 odd}\,=\,20.6\,\bas^3\,mb \,\approx\,0.5 \,mb $ for $\bas = 0.3$ in \eq{ODD}
 we take from the QCD estimates in Ref.\cite{RYODD}.
 The value of $B_{\rm odd}=5.6\,GeV^{-2}$ which is smaller than elastic slope
 for the BFKL Pomeron in accord with QCD estimates\cite{RYODD}. The sign minus
 in \eq{ODD} corresponds to proton-proton scattering, while the sign plus
 refers to
  antiproton-proton collisions. Our odderon parameters are in
 accord with the estimates in Ref.\cite{KMR}. The amplitude $f(s,t)$ is
 related to $a_{el}\Lb s,b\Rb$ by \eq{OBELT} (see also \eq{OBS1}).
 
 In \fig{dsdt2} we show the prediction for proton-antiproton scattering. 
One can conclude that in our model the measurements of the elastic cross
 sections for $p\,p$ and $\bar{p} p$ scattering can provide the estimates 
for
 the odderon contribution. It should be stressed that the contribution
 of the $\omega$-reggeon leads to negligible contribution at
 $W= 7\,TeV$ (see \fig{dsdt1}).

       \begin{figure}[ht]
    \centering
  \leavevmode
      \includegraphics[width=10cm]{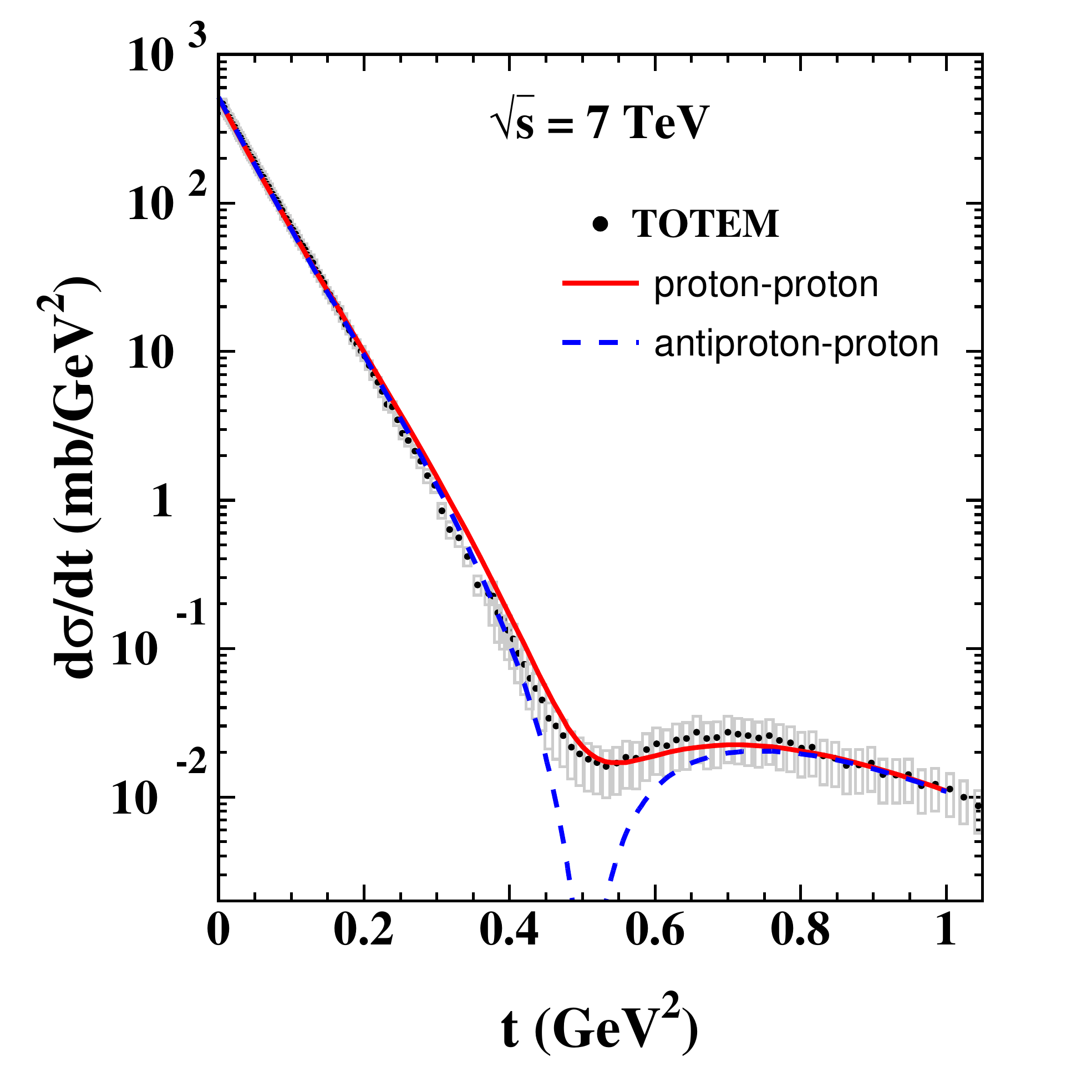}       
             \caption{$d \sigma_{el}/dt$ versus $t$. 
 The solid line describes the elastic cross sections
 for $p p$-scattering  with the odderon contribution
 (see \eq{ODD}), while the dashed line shows the 
elastic cross section for $\bar{p} p$-scattering 
using \eq{ODD}. The data are taken from Ref.\cite{TOTEMLT})
              }      
\label{dsdt2} 
   \end{figure}


~

~


 \section{Conclusions}
The primary goal of this paper was to investigate whether the new
 parton model, which has been developed in Ref.\cite{GLPPM}, is able
 to describe the diffraction production.
 The model is
 based on  Pomeron calculus in 1+1 space-time, suggested in Ref.
 \cite{AKLL}, and on the simple assumptions on the hadron structure,
 related to the impact parameter dependence of the scattering amplitude.
 This parton model stems from QCD, assuming that the unknown non-perturbative
 corrections lead to fixing  the size of the interacting dipoles.
 The advantage of this approach is that it satisfies both  t-channel
 and s-channel unitarity, and it  can be used for summing all diagrams of
 the Pomeron interaction including the Pomeron loops.  Our hope was that
 this model will be  superior to  the model which we developed 
based on
 CGC approach\cite{GLP1,GLP2},  and which does not satisfy both $t$ and
 $s$ channel unitarity.
 
 Unfortunately, we did not find  any advantages of our new model, and we 
have
 to describe half of the single diffraction cross section by the diffraction
 production of large masses, in striking similarity with the CGC based 
models.
 Certainly, it is not a very encouraging result, especially  since
the CGC 
models describe the large mass diffraction production better than this
 model. Mostly  this is due to the fact that $\Delta_{\rm dressed}$ 
in this model,
 turns out to be larger than in CGC one.

The impact parameter dependance of the scattering amplitudes (see \fig{ael})
 shows that the soft interaction at high energies measured at the LHC have
 a much richer structure that we  presumed in the past. We believe 
that we  have demonstrated
 that the character of high energy scattering is closely related to the
 structure of hadron,  which presently is described by 
a simple two
 channel model.
 
  Our attempt to describe the $t$-dependence of the elastic
 cross section shows that we can reproduce the main features of
 the $t$-dependence that are measured experimentally: the slope
 of the elastic cross section at small $t$, the existence of the
 minima in $t$-dependence which is located at $|t|_{min} = 0.52\,GeV^2$
 at W= 7 TeV; and the behaviour of the cross section at $|t|\,>\,|t|_{min}$.
 It should be stressed  that our model allows us to find the real part
 of the scattering amplitude using our general expression of \eq{AIK} 
 for $A_{i 
k}\Lb s,t\Rb$.  We consider the sum $A_{ik}\Lb s,+ i \epsilon t\Rb +
 A_{ik}\Lb u-i \epsilon,t\Rb$, which corresponds to positive signature,
 and calculated the real part of this sum. 
 It should be stressed that
 we do not use  any of  the simplified approaches to estimate the 
real part of
 the amplitude which we show (in our model ) which  do not reproduce
 correctly the real part of the amplitude at large $t$.
  In our model the real
 part turns out to be much smaller 
 than the experimental one. Consequently,  to achieve a description of the 
data, it is necessary  to add an 
odderon
 contribution.  Hence, our model
 corroborates the conclusion of Ref.\cite{ODDSC}.

  A topic for future study, is whether the characteristic behaviour
 of the $A_{i,k}(b)$ amplitudes as a function of $b$ stems from the theory of
 interacting Pomerons, 
 which
   satisfies both $s$ and $t$ channel unitarity, or  is an
 artifact of the simple two channel approach with the phenomenological 
input,
 on the impact parameter dependence.

We are  aware that our model is very naive in describing  the
 hadron structure, but hope that   further progress in accumulating
 data on diffraction production, as well as the unsolved problem  of 
 treating  the processes of the multiparticle generation in the
 framework of our approach,  will generate a self consistent
 picture for  high energy scattering at long distances.

 {\it Acknowledgements.} \\
   We thank our colleagues at Tel Aviv University and UTFSM for
 encouraging discussions.  Our special thanks go to
 Tamas Cs\"org\H o and Jan Kasper for discussion of the odderon contribution
 and elastic scattering   during the Low x'2019 WS. 
 This research was supported  by 
 CONICYT PIA/BASAL FB0821(Chile)  and  Fondecyt (Chile) grants  
1170319 and 1180118 .

    \end{document}